\documentclass[usenatbib]{mn2e}
\usepackage{graphicx}
\usepackage{natbib}
\usepackage{amssymb}
\usepackage{aas_macros}
\def\degg{\hbox{$\null^\circ$\hskip-3pt.}}

\title[Distances and Metallicities in  the Local Group] {Distances and
Metallicities  for 17  Local Group  Galaxies} \author  [McConnachie et
al.]   {A.  W.  McConnachie${^1}$,  M.  J.   Irwin${^1}$,  A.  M.   N.
Ferguson${^2}$, R.  A.  Ibata${^3}$,\newauthor G.  F.  Lewis${^4}$, N.
Tanvir${^5}$\\   ${^1}$  Institute   of  Astronomy,   Madingley  Road,
Cambridge,   CB3  0HA,   U.K.\\  ${^2}$   Max-Planck-Institut  f\"{u}r
Astrophysik,   Karl-Schwarzschild-Str.  1,   Postfach   1317,  D-85741
Garching,  Germany\\ ${^3}$  Observatoire  de Strasbourg,  11, rue  de
l'Universite,  F-67000,  Strasbourg,   France\\  ${^4}$  Institute  of
Astronomy,  School of Physics,  A29, University  of Sydney,  NSW 2006,
Australia\\  ${^5}$   Physical  Sciences,  Univ.    of  Hertfordshire,
Hatfield, AL10 9AB, U.K.\\}

\begin{document}

\maketitle

\begin{abstract}
We have obtained Johnson V and Gunn i photometry for a large number of
Local Group galaxies using the Isaac Newton Telescope Wide Field
Camera (INT~WFC).  The majority of these galaxies are members of the
M31 subgroup and the observations are deep enough to study the top few
magnitudes of the red giant branch in each system.  We previously
measured the location of the tip of the red giant branch (TRGB) for
Andromeda~I, Andromeda~II and M33 to within systematic uncertainties
of typically $< 0.05$~mags (\citealt{mcconnachie2004a}).  As the TRGB
acts as a standard candle in old, metal poor stellar populations, we
were able to derive distances to each of these galaxies.  Here we
derive TRGB distances to the giant spiral galaxy M31 and 13 additional
dwarf galaxies - NGC~205, NGC~185, NGC~147, Pegasus, WLM, LGS3, Cetus,
Aquarius, And~III, And~V, And~VI, And~VII and the newly discovered
dwarf spheroidal And~IX.  The observations for each of the dwarf
galaxies were intentionally taken in photometric conditions.  In
addition to the distances, we also self-consistently derive the median
metallicity of each system from the colour of their red giant
branches. This allows us to take into account the small metallicity
variation of the absolute I magnitude of the TRGB.  The homogeneous
nature of our data and the identical analysis applied to each of the
17 Local Group galaxies ensures that these estimates form a reliable
set of distance and metallicity determinations that are ideal for
comparative studies of Local Group galaxy properties.
\end{abstract}


\section{Introduction}

The Local Group is home to some 35 galaxies, the majority of which are
suspected to be satellites of  the two most massive bodies, the Galaxy
and         M31        (for         recent         reviews,        see
\citealt{mateo1998a,lyndenbell1999,vandenbergh1999}).    Many  of  the
dwarfs, such as M32, have  been known about for decades, while others,
such   as   And   V,   VI,   VII  \&   IX   are   recent   discoveries
(\citealt{armandroff1998,armandroff1999,karachentsev1999,zucker2004}). As
the  discovery of  And  IX by  \cite{zucker2004}  demonstrates, it  is
unlikely that  even now we have  a full inventory of  the Local Group,
and no  doubt some  faint bodies still  await discovery.   Interest in
Local  Group galaxies  stems from  their proximity  and representative
nature, since they are relatively easily resolved and include examples
over a wide range of luminosities.

With so many galaxies known in  the Local Group it has become possible
to conduct comparative studies  of the different population types such
as  the dwarf  irregulars  and dwarf  spheroidals.  However,  accurate
studies  of stellar  populations  requires accurate  knowledge of  the
distance to each galaxy, and this becomes significantly more important
when  populations  in different  galaxies  are  to  be compared.   The
distances to the nearby satellites of the Milky Way are believed to be
known to  reasonable accuracy and are  based on a  variety of methods.
However, this is not always the  case for the more distant Local Group
galaxies, such as the satellites of M31.  Furthermore, the majority of
distance  determinations  to these  galaxies  have  been conducted  as
independent   studies  by   different  groups.   It   is  nevertheless
advantageous to acquire a  homogeneous set of distance determinations.
This  then removes  systematic  uncertainties that  may exist  between
measurements  that  could  be  due to  the  data  aquisition/reduction
process,  the standard  candle  employed, and  the  algorithm that  is
applied.   Reliable relative  distances to  the galaxies  can  then be
obtained,  and  provides  much  of  the  motivation  behind  the  work
presented in this paper.

All of  the Local Group  galaxies contain a significant  population of
old,  relatively metal-poor  stars (Population  II). For  M31  and the
Galaxy, these stars are found predominantly in the stellar halo, while
for the dwarfs, Population~II stars tend to be the dominant component.
A lack of Cepheid variables in  this population means that in order to
estimate  distances to  them other  indicators have  to be  used.  One
obvious  candidate  is  based on  the  Tip  of  the Red  Giant  Branch
(TRGB). This point in stellar evolution marks the onset of core helium
burning in  Red Giant  Branch (RGB) stars  and observation  and theory
show  it to  occur  at  a relatively  constant  I-band magnitude  (eg.
\citealt{dacosta1990,salaris1997}).   \cite{barker2004}  have recently
conducted  a  detailed study  of  the reliability  of  the  TRGB as  a
distance  indicator for  stellar populations  with a  variety  of star
formation histories and find that  it is reliable for any system which
is not  dominated by  a metal-rich population  ([Fe/H]$ > -0.3$)  or a
population   with   a    substantial   young   ($\lesssim   1.7$~Gyrs)
component. This makes it ideal  for calculating distances in the Local
Group.

\cite{lee1993a}   were  influential   in  developing   a  quantitative
technique, based upon a  Sobel edge detection algorithm, for detecting
the  location of  the TRGB  in a  photometric dataset.   It  was later
adapted   and  refined   by  \cite{sakai1996}   and,   more  recently,
\cite{mendez2002} who also developed  a maximum likelihood approach to
estimate  the TRGB  location.  Edge  detection  algorithms essentially
assume  that the  luminosity function  in the  region of  the  tip are
step-like,  and  their  simplest  application looks  for  the  largest
absolute  change  in star  counts  between  neighbouring  bins in  the
luminosity  function.  As  such, they  can be  easily affected  by the
noise features which are often prevalent in real luminosity functions
on which TRGB analyses are to be conducted.

In a  previous paper (\citealt{mcconnachie2004a},  hereafter Paper~I),
we developed a  new technique to allow for  the accurate determination
of the TRGB as represented  in a photometric dataset. We advocated the
use  of Wide  Field Cameras  for such  a study  as this  maximises the
number of stars that contribute to the luminosity function, decreasing
the effects  of poisson noise on  the measurement and  ensuring that a
fair sample of the brightest RGB stars are present. In Paper~I we also
introduced,  tested   and  implemented   a  new  algorithm   for  TRGB
determination which uses a  data-adaptive slope to fit the `luminosity
probability distribution'  (LPD) in  the region of  the tip  and hence
derive a distance modulus for the system. The LPD is given by

\begin{equation}
\label{smooth}
\wp\left(m\right)=\sum_{i=1}^{N_\star} \frac{1}{\sqrt{2\pi}\sigma_i}\exp\left(-\frac{\left(m_i-m\right)^2}{2\sigma_i}\right),
\end{equation}

\noindent where $m_i$ and $\sigma_i$ are the magnitude and photometric
error of the $i^{th}$ star  from a sample of $N_\star$. This technique
was shown  to have systematic uncertainties usually  of order \mbox{$<
0.05$~mags.}  We went  on to calculate the distances,  as derived from
data  taken  with  the   Isaac  Newton  Telescope  Wide  Field  Camera
(INT~WFC), to  the dwarf galaxies And~I  \& II, and  the spiral galaxy
M33.  The  reader is referred  to this paper, and  references therein,
for a discussion of the practicalities involved with this technique.

In this paper, we significantly extend the number of Local Group
galaxies for which we have measured locations of the TRGB and
distances, based on a large homogeneous photometric database of Local
Group galaxies. This data has been taken with the INT WFC, a four-chip
EEV 4K x 2K CCD mosaic camera with a $\sim 0.29 \square^\circ$ field
of view (\citealt{walton2001}).  Over the past three years we have
been using this instrument in conjunction with the
Canada-France-Hawaii 12K Camera to conduct a photometric survey of M31
(\citealt{ibata2001a,ferguson2002,mcconnachie2003}, Irwin et al., {\it
in preparation}), which is now complete over an area of 40
$\square^\circ$.  We have also presented photometry for M33, And~I,
And~II (Paper I; see also Ferguson et al., {\it in preparation}), and
here we present colour magnitude diagrams (CMDs), metallicity
information and distance determinations to And~III, And~V, And~VI,
And~VII, And~IX, LGS3, Cetus, WLM, Pegasus, Aquarius, NGC~147,
NGC~185, NGC~205 and M31. These galaxies constitute the majority of
Local Group members visible from the northern hemisphere.  M32 is not
included in this list due to severe crowding problems and its awkward
projected location close to the centre of M31. Our data for IC10 is
likewise not presented here as this object's close proximity to the
Galactic plane causes a serious amount of differential reddening which
complicates its analysis.  For the rest, the resulting set of
homogeneous distance estimates is the largest that has been measured
for the Local Group: the data for each galaxy were taken using the
same telescope and the same instrument, through the same filters for
similar exposure times.

The data for the dwarf galaxies was taken in conjunction with the main
M31  and  M33  surveys.   Observations  for  each  dwarf  galaxy  were
restricted to  photometric conditions simplfying  cross-calibration of
all the data to the same photometric system.  All data taken were then
reduced and calibrated in exactly  the same way with the same pipeline
processing  (\citealt{irwin2001}).  The  subsequent analysis  for each
galaxy was also identical, with  the same algorithm applied to each to
derive  the  distance  estimates.   As such,  differential  systematic
errors  are  minimised  and  the  entire dataset  is  thus  ideal  for
comparative studies of these galaxies.

Section 2 provides an overview of our photometry, discusses the
reddening estimates, the assumed absolute I magnitude of the TRGB, the
metallicity determination and how we deal with foreground and
background contamination.  In Section~3 we briefly review the current
knowledge of the stellar content of each of the galaxies and calculate
the median metallicity and distance to each from the INT WFC data.  We
finish in Section~4 with a discussion of our results and a comparison
to previous work.
 
\section{Preliminaries}

\subsection{Photometry}

All the photometry for the dwarf galaxies analysed in this paper
consist of a single targetted WFC pointing, with the exception of And~III, which fell across two pointings, as part of the M31 survey.  The
data for the isolated dwarf galaxies were all systematically taken in
photometric conditions as part of two separate M31 and M33 survey runs
in August and September 2003.  As for the main surveys, the INT~WFC
Gunn i (${\rm i'}$) and Johnson V (${\rm V'}$) passbands were used.
Exposure times were 800-1000 s per passband and allowed us to probe
down to magnitudes of 23.5\,mags in ${\rm i'}$ and 24.5\,mags in ${\rm
V'}$ (S/N $\simeq$ 5).  Depending upon the galaxy, this is usually
sufficient to detect individual RGB stars to an absolute magnitude of
${\rm V'} \simeq 0$ and main-sequence stars to ${\rm V'} \simeq -1$.
For the subsequent analysis, we have converted our data to the Landolt
equivalent passbands ${\rm V}$ and ${\rm I}$ (\citealt{landolt1992}).
The transformations required are ${\rm I} ={\rm i'} - 0.101 \times
{\rm (V-I)}$ and ${\rm V} = {\rm V'} + 0.005 \times {\rm (V-I)}$.
These transformations have been derived by comparison with
observations of several Landolt standard
fields\footnote{http://www.ast.cam.ac.uk/$\sim$wfcsur/colours.php}. The
standard INT Wide Field Survey (WFS) pipeline, supplied by the
Cambridge Astronomical Survey Unit (\citealt{irwin2001}), was used to
process all of the on-target data plus calibration frames, as was the
case for Paper I.  Our final uncertainty in the distances we derive
includes the average $rms$ photometric error in the region of the
TRGB, which is typically 0.02~mags at ${\rm I} \simeq 20.5$~mags, and
the average photometric zeropoint calibration error, also of order
0.02~mags.

Objects are classified independently in each passband based on their
overall morphological properties, specifically their ellipicity as
derived from intensity-weighted second moments and the curve-of-growth
of their flux distribution (\citealt{irwin2004}).  Measures from these
are combined to produce a normalised N(0,1) statistic which compares
the likeness of each object to the well-defined stellar locus visible
on each frame. Stellar objects are chosen to lie within 2 or $3\,-\,\sigma$ of this locus depending on the desired tradeoff between
completeness and contamination from non-stellar objects (usually
compact galaxies or spurious images).  For the purposes of this study,
we generally make use of all objects which lie within $3\,-\,\sigma$ of
the stellar locus.  For situations in which many sources are available
to us, we use the more stringent $2\,-\,\sigma$ limit.  At faint
magnitudes, typically within \mbox{1 - 2 mags} of the frame limit
(depending on seeing), the stellar locus is vulnerable to
contamination by distant compact galaxies.  However, this has little
effect here as we are mainly concerned with the relatively bright
stars near the TRGB.

\subsection{Foreground contamination}

Each red giant branch luminosity function/LPD is statistically
corrected for foreground stars, as was the case in Paper~I. Except for
M31 and NGC~205, this is done in a standard way. In constructing the
CMD for the galaxy, we try to use stars within $\sim 0.2^\circ$ of the
centre of the galaxy ie. approximately 40\% of the area of the INT WFC
field is used.  Allowing for foreground stars, this means that we
generally use somewhere in the region of 50 - 100\,\% of the stars
that belong to the dwarf galaxy that we have detected.  For some of
the smaller, fainter objects, such as And~V or Aquarius, we have been
forced to use a smaller area to more clearly detect the RGB over and
above the Galactic foreground. However, even in these situations we
are still sampling all the bright RGB stars within $\sim 5$ core radii
($r_c$).  Generally, the outer regions of each INT~WFC pointing are
used in constructing a reference luminosity function.  This is scaled
to the actual luminosity function either by matching the number of
brighter foreground stars in each zone or by simply computing the area
ratio used.  In the case of M31 and NGC~205 we adopted a different
strategy making use of the larger area survey available; details of
this are given in the appropriate section.  Additionally, the TRGB
location in M33 has been rederived making use of a local foreground
correction in exactly the same way as will be described for M31, and
using only objects lying within $2\,-\,\sigma$ of the stellar
locus. The former alteration causes a $0.03$\,mags difference with the
result presented in Paper~I (see Tables 1~\&~2).

\subsection{Reddening corrections}

All reddening corrections used in this paper are taken from
\cite{schlegel1998}.  These authors conducted an all-sky survey of
infra-red dust emission and used this to calculate the reddening,
${\rm E\left(B - V\right)}$, across the sky to within an uncertainty
of $\sim 16\%$.  We have also included the effects of this error in
our final error estimation. The extinction in the I-band is calculated
by the relationship given in the same paper, \mbox{${\rm A_I} =
1.94\,{\rm E\left(B - V\right)}$.}  We note that \cite{arce1999} and
\cite{bonifacio2000} have suggested that the Schlegel maps may
overestimate the reddenning values in areas where the colour excess is
larger than \mbox{E(B - V) $\simeq 0.1$\,mags}.  We do not consider
this effect here for the sake of homogeneity, and if real it will
affect very few of our results, none of them significantly \mbox{($<
0.035$\,mags)}. All adopted reddenning values are listed in Table~1 to
allows the distance modulii to be recalculated should later work
produce revised estimates for the reddening.

\subsection{Absolute I magnitude of the TRGB}

In Paper~I, we adopted the blanket value of $4.04 \pm 0.05$ mags as
the absolute I magnitude of the TRGB, based upon the value given by
\cite{bellazzini2001}.  These authors calculated $M_I^{TRGB} = -4.04
\pm 0.12$\,mags at \mbox{${\rm [Fe/H]} \sim -1.7$} by analysis of the
red giant branch of Omega Centauri. The large uncertainty in this
value comes primarily from the uncertainty in the distance modulus to
$\omega$\,Cen.  The value they adopted for this was derived by
\cite{thompson2001} by observations of a detached eclipsing binary,
for which a value of $\left(m - M\right)_o = 13.65 \pm 0.11$ was
calculated.  The value of $M_I^{TRGB}$ calculated by
\cite{bellazzini2001} is in excellent agreement with that derived by
other means, and the quoted uncertainty is somewhat conservative.  For
this reason we adopted the smaller uncertainty of 0.05~mags, but noted
that it is possible for the entire scale to shift by some larger
amount, affecting all our distance estimates in the same fashion.

In this paper, we now also take into account the slight variation that
$M_I^{TRGB}$        displays        with       metallicity        (eg.
\citealt{bellazzini2001,bellazzini2004}).   This is generally  a small
effect; however, the spread in  metallicity shown by the galaxies that
we are analysing  necessitates that we take this  effect into account.
The   most    recent   empirical    calibration   of   this    is   by
\cite{bellazzini2004} who fit the following relationship;

\begin{eqnarray}
M_I^{TRGB} &=& 0.258 \left[\frac{M}{H}\right]^2 + 0.676 \left[\frac{M}{H}\right] - 3.629 \nonumber \\
&=& f\left(\left[\frac{M}{H}\right]\right)
\end{eqnarray}

\noindent            At           intermediate           metallicities
\mbox{($\left[\frac{M}{H}\right]  \sim -1.5$)} this  relation predicts
values  of  $M_I^{TRGB}$ similar  to  the  `standard'  value of  $\sim
-4.04$~mags.  However, at very  low metallicities (eg. LGS3, Aquarius;
$\left[\frac{M}{H}\right]   <   -2$)    the   values   predicted   are
substantially  fainter ($M_I^{TRGB}  \sim  -3.95$~mags) than  normally
assumed.  It  is unclear  at this time  whether such faint  values are
realistic or what the  physical explanation for the reduced luminosity
would be.  Although line blanketing effects lead to a reduction in the
I~band  luminosity at high  metallicity, an equivalent  dimming effect
has not been reported in  the literature for low metallicity.  Our own
work  on M31  suggests there  is no  strong evidence  to  suggest that
$M_I^{TRGB}$ becomes fainter  at low metallicity and we  find that the
behaviour of $M_I^{TRGB}$ with metallicity is well approximated by the
evolutionary          tracks         of         \cite{dvandenberg2000}
(\citealt{mcconnachie2003}).   Since  the   dimming  may  well  be  an
artefact  of the empirical  calibration used,  for systems  more metal
poor than  $\omega$\,Cen, we adopt \mbox{$M_I^{TRGB}  = {\rm min}\left[-4.04,
f\left(\left[\frac{M}{H}\right]\right)\right]$.} Although this modified
calibration is still {\it  adhoc}, it nevertheless prevents physically
implausible levels of dimming at low metallicity.

In  Tables~1~\&~2  we  provide the  reader  with  all the  neccessary
information  to recalculate  the  distance modulii  should later  work
produce   improved  calibrations.    The  distances   to  Andromeda~I,
Andromeda~II  and M33,  which were  originally calculated  in Paper~I,
have  been   recalculated  to  account  for   this  small  metallicity
dependancy.   The  results  for  these  galaxies are  also  listed  in
Tables~1~\&~2 along with the 14 galaxies examined here.

\subsubsection{Calculating $\left[\frac{M}{H}\right]$}

In order to maintain internal consistency and also as an independent
metallicity measure, we choose not to use metallicity estimates for
each system from the literature but instead calculate the metallicity
of each system from the locus of the RGB in the INT~WFC data.
\cite{dvandenberg2000} produced an extensive set of theoretical
evolutionary tracks that trace stellar evolution up to the point of
core helium ignition.  A metallicity distribution function (MDF) can
be readily constructed for each system using these tracks by making
the plausible assumptions that the majority population seen on the
putative RGB is composed of first ascent giants, and has an age
greater than 2~Gyrs.  A full description of the technique used to
construct the MDFs and an analysis of the metallicity information for
each galaxy will be presented elsewhere (McConnachie, {\it in
preparation}) together with a detailed study of the metallicity
variation in and around M31.  It should be noted that our technique is
essentially the same as that employed by numerous other authors (see,
for example \citealt{durrell2001,durrell2004,bellazzini2003}) and so
we give only a brief description below.

The evolutionary tracks are first shifted in colour-magnitude space so
that they lie at the reddening and (approximate) distance of the
system we wish to analyse. For each star in the top two magnitudes of
the RGB we then calculate its metallicity using a bilinear
interpolation from the evolutionary tracks that lie either side of the
star.  A MDF is then obtained by constructing a histogram of the
resulting metallicity estimates.  The foreground is removed by
constructing a suitable reference field and subtracting it from the
galaxy MDF.  However, since $M_I^{TRGB}$ also depends on metallicity
then so does the distance against which the tracks are registered. We
thus need to iterate towards a final solution for the metallicity and
distance modulus using the newly estimated median metallicity for each
system, to update the value of $M_I^{TRGB}$ and then repeating the
procedure until it converges. The accuracy of this technique is
estimated to be $\pm\,0.1\,-\,0.2$\,dex.

This process is conducted twice for each system, using tracks with no
$\alpha$-element enhancement and tracks with an $\alpha$-element
enhancement of 0.3.  Although there is little evidence to suggest that
any of the dwarfs have an $\alpha$ enhancement, we do not want to
misinterprete a lack of evidence as evidence for a lack. It transpires
that the median value of $\left[\frac{M}{H}\right]$ is relatively
insensitive to the $\alpha$-element abundance assumed, and the implied
value of $M_I^{TRGB}$ changes by, at most, a few hundredths of a
magnitude.  This has a negligible effect on the distance that we would
calculate and we are thus able to derive a unique distance to each
galaxy.

Although this procedure was adopted for the majority of the galaxies
studied, the much broader metallicity spread for M31 and NGC~205 (and
M33) required a slight variant on this scheme.  Generally, the strip
of stars used in calculating the position of the TRGB spans
approximately the entire RGB, and thus the median metallicity of the
RGB is essentially the representative metallicity of the tip.  For the
three cases mentioned here, however, the strip used to calculate the
position of the tip is substantially narrower than the RGB, and so the
representative metallicity of the tip is {\it not} well approximated
by the median metallicity of the RGB. In these cases, we instead use
the metallicity of the evolutionary track that best matches the
position of the chosen analysis strip.  Without this restriction the
much larger spread in metallicities on the metal-rich side for these
galaxies blurs out the position of the tip and also systematically
lowers $M_I^{TRGB}$.  Furthermore, by restricting the range in
metallicity used we also reduce the dependence on the empirical
relationship in Equation~2.  For completeness, we still include the
calculated RGB median metallicities for these systems in Table~1.

\section{TRGB Distance Determinations}

Figures~1~-~14 show the CMDs for each galaxy (left panel).  The
luminosity fuction (histogram in upper right panel) and the LPD on
which our algorithm is applied (offset curve in upper right panel) are
constructed only from stars which are limited by the dashed lines in
the CMD, to reduce contamination from other stellar populations. An
arrow shows the measured position of the TRGB on the luminosity
functions, and a horizontal dashed line indicates this location on the
CMD. This line does not extend for the full width of the CMD to
allow the reader to make an independent judgement of the location of
the TRGB.

The output of our secondary, heuristic TRGB-finding technique is shown
in the lower right panel. This method involves taking the ratio of the
star counts  in neighbouring bins  from the luminosity  histogram once
the counts have  passed a threshold level (see  Paper~I).  Each bin is
averaged with  its immediate neighbours  to attempt to  reduce poisson
noise  effects.   This technique  provides  a  `sanity  check' of  our
data-adaptive slope  technique and  should be in  reasonable agreement
with the  more involved method.   However, this technique has  all the
usual  problems associated  with  edge-detection algorithms  discussed
previously, albeit this time in the log domain. This, coupled with the
local neighbourhood averaging, leads  to a tendency for this heuristic
method  to   peak  at  systematically  fainter   magnitudes  than  the
data-adaptive slope  method, although the simplicity  of the technique
makes it attractive as a first pass on the data.

\subsection{M31, the Andromeda Galaxy}

M31 ($0{\rm  h}~42{\rm m} ~44.3{\rm  s}, +41^\circ ~16' ~9''$)  is the
closest giant  spiral galaxy  to the Milky  Way. It provides  a unique
opportunity to study in detail a  galaxy that is thought to be similar
to our own. All the stellar  populations that one would expect to find
in a large disk galaxy can  be found in M31, and a detailed discussion
of this vast  and spectacular object is well beyond  the scope of this
paper.

\cite{ferguson2002} and Irwin et al., {\it in preparation}, present
maps of the spatial distribution of red giant branch stars as seen in
our INT~WFC survey of this galaxy.  Large amounts of substructure are
obvious. However, we need to ensure that our distance estimate to this
galaxy is unbiased by these features. For this reason, our analysis
uses only those stars that are located within an elliptical annulus, $e = 0.4$, centred on M31, with a semi-major axis ranging from
2\degg25 to 2\degg5 ($\sim 35$ to $40$\,kpc). This `halo' zone is far
enough from the centre of M31 to be not significantly affected by
crowding or serious contamination from disk and/or bulge components.
As the substructure revealed in our survey is spatially distinct its
contribution within this annulus is small compared to the generic M31
halo population.  We use as a reference field stars located in a
similar, more distant, elliptical annulus with a semi-major axis
ranging from 2\degg5 to 2\degg8.

\begin{figure*}
\begin{center}
\includegraphics[angle=270, width=13.5cm]{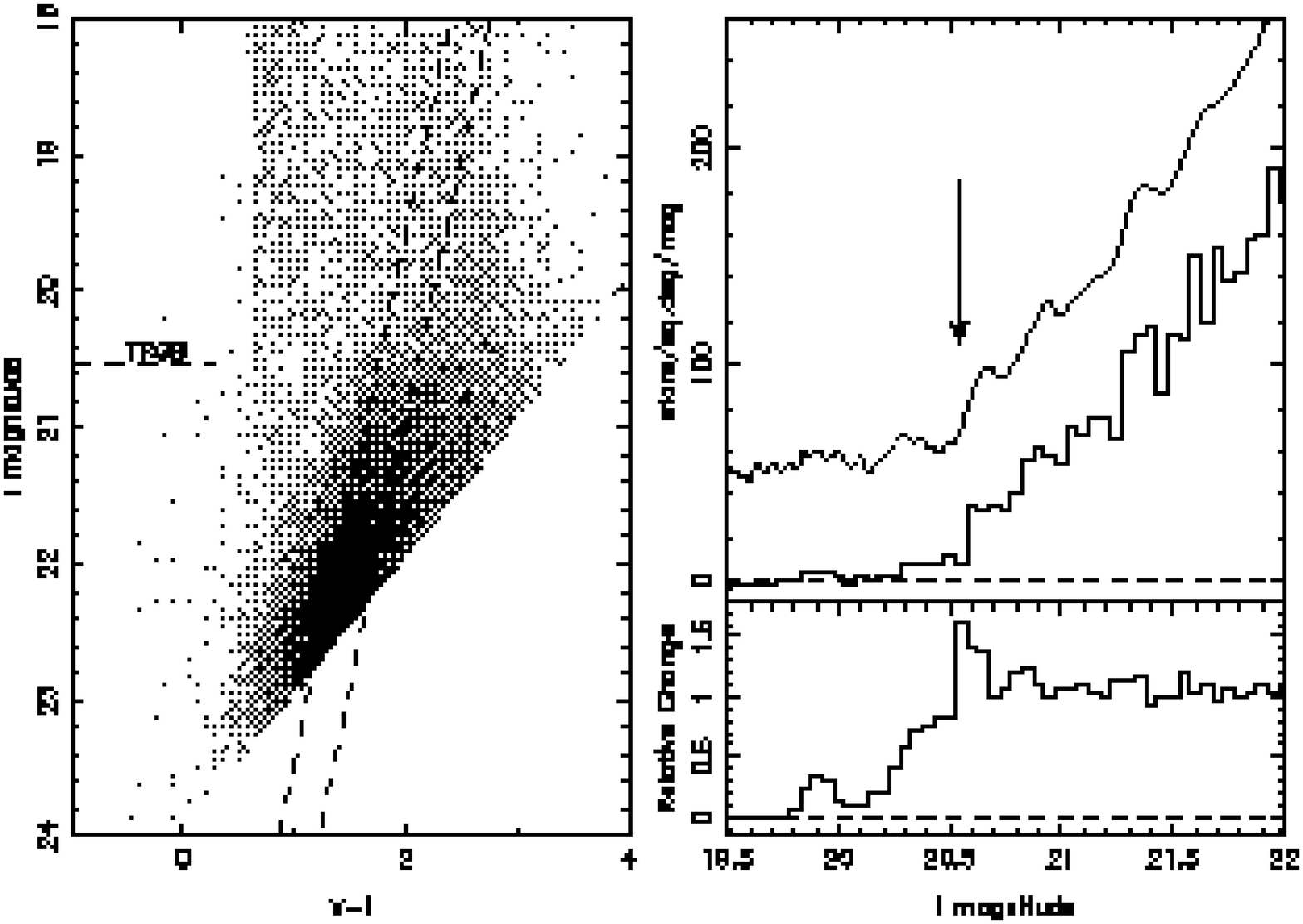}
\caption{Hess  diagram with  square-root scaling  showing the  M31 CMD
(left  panel), RGB  luminosity function  and offset  LPD  (upper right
panel),  and  heuristic  signal  (lower  right panel).   The  TRGB  is
measured to lie at ${\rm I} =  20.54$~mags and is marked on the CMD by
a horizontal dashed line and  on the luminosity functions by an arrow:
stars within an elliptical annulus of $e = 0.4$ centred on M31, with a
semi-major axis  ranging from  2\degg25 to 2\degg5,  were used  in our
analysis. Stars in an annulus outside of this were used as a reference
field to correct for the foreground population.}
\label{m31}
\end{center}
\end{figure*}

The CMD and RGB luminosity functions for M31 are shown in
Figure~\ref{m31}.  These are constructed from the elliptical annuli
described above using all objects that lie within $3\,-\,\sigma$ of
the stellar locus, maximising the number of stars available to us at
this galactocentric radius.  The TRGB is very obvious in these
diagrams, and the TRGB algorithm applied to the LPD gives the results
detailed below. Note that the metallicity used to calculate
$M_I^{TRGB}$ for this galaxy is not the median metallicity of the RGB
but the representative metallicity of the strip of stars shown in the
left panel of Figure~\ref{m31} \mbox{([M/H] $\sim -1$).}  Discussion
of this result, and the following results, will be deferred until
Section~4.

\begin{eqnarray*}
M31:\\
{\rm I_{TRGB}}           & = & 20.54 \pm 0.03~{\rm mags} \nonumber\\
{\rm E\left(B - V\right)}& = & 0.06~{\rm mags}        \nonumber\\
{\rm \left[M/H\right]_{\alpha=0.0}}&=&-0.6\nonumber\\
{\rm \left[M/H\right]_{\alpha=0.3}}&=&-0.5\nonumber\\
{\rm M_I^{TRGB}}         & = & -4.05~{\rm mags}     \nonumber\\
{\rm \left(m - M\right)_o} & = & 24.47 \pm 0.07~{\rm mags}\nonumber\\
{\rm D_{M31}}        & = & 785 \pm 25~{\rm kpc}      \nonumber\\
\\
M31~Error~Budget: \nonumber \\
{\rm Photometry}-~~~rms   & : & \pm 0.03~{\rm mags} \nonumber\\
{\rm ~~~~~~~~~~}~-zeropt   & : & \pm 0.02~{\rm mags} \nonumber\\
{\rm Reddening}    & : & \pm 0.02~{\rm mags} \nonumber\\
{\rm M_I^{TRGB}}   & : &  \pm 0.05~{\rm mags} \nonumber\\
{\rm Algorithm}    & : &  \pm 0.03~{\rm mags} \nonumber\\
{\rm Total}        & : &  \pm 0.07~{\rm mags} \nonumber\\
\end{eqnarray*}

\subsection{NGC 205}

NGC  205 ($0{\rm  h}~40{\rm m}  ~22.1{\rm s},  +41^\circ  ~41' ~7''$),
along  with NGC  185 and  NGC 147,  is one  of three  dwarf elliptical
companions  to M31.   It lies  only  $37^\prime$ ($\simeq  9$ kpc)  in
projection from Andromeda and shows evidence of tidal interaction with
its                massive                companion               (eg.
\citealt{hodge1973,choi2002,mcconnachie2004b}).            Multi-colour
photometry of this galaxy's inner regions by \cite{lee1996} revealed a
blue-plume of stars and a  bright AGB population, indicating that both
a young and intermediate age  population are present.  Earlier work by
\cite{mould1984}  had shown that  its metallicity  was of  order ${\rm
[Fe/H]} \simeq -0.9$, with a dispersion of $\sim 0.5$.

\begin{figure*}
\begin{center}
\includegraphics[angle=270, width=13.5cm]{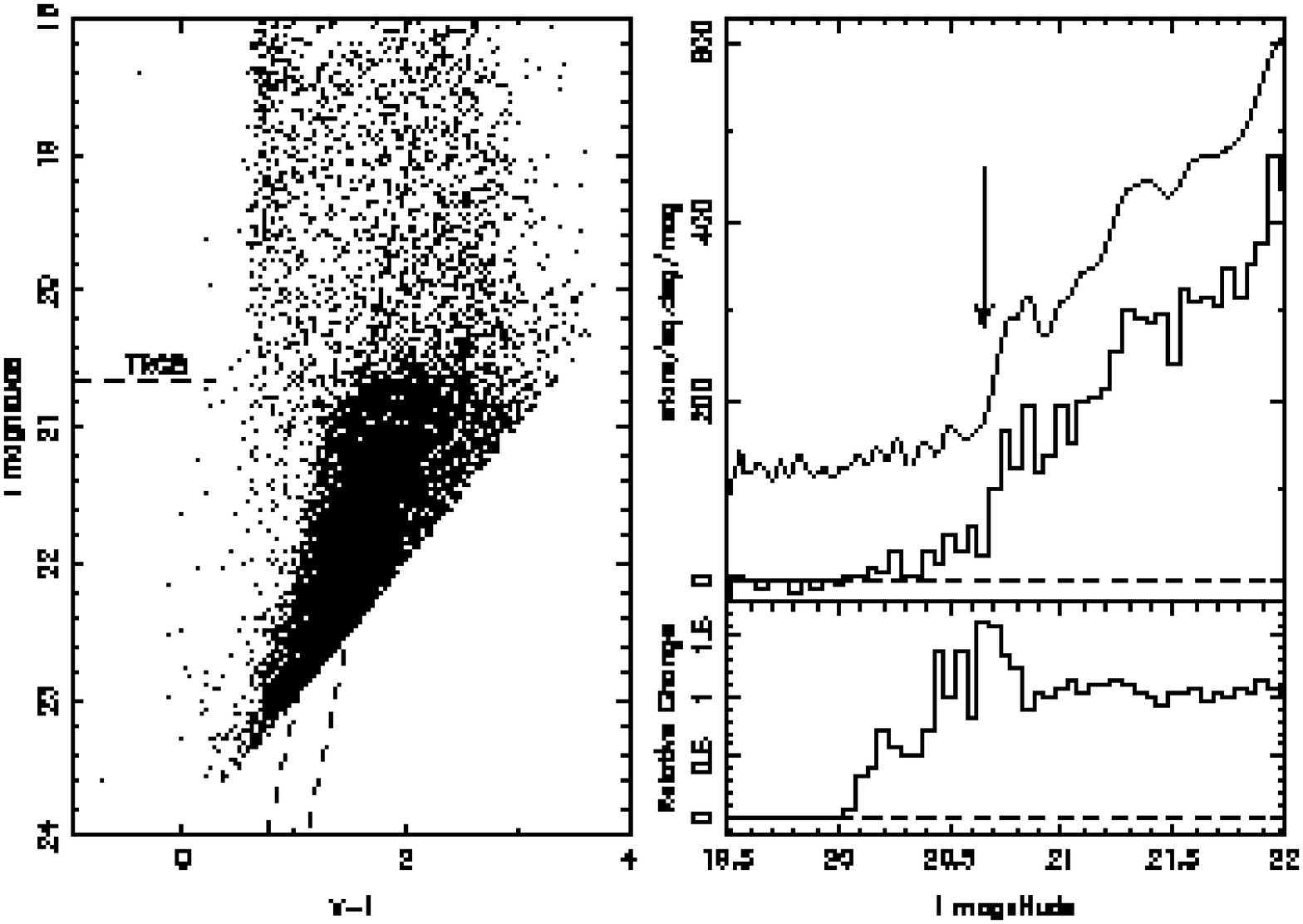}
\caption{NGC~205 CMD (left panel),  RGB luminosity function and offset
LPD (upper right panel), and heuristic signal (lower right panel). The
TRGB is measured to lie at ${\rm I} = 20.65$~mags and is marked on the
CMD by a horizontal dashed line  and on the luminosity functions by an
arrow.}
\label{ngc205}
\end{center}
\end{figure*}

NGC~205 falls  well within  the area surveyed  as part of  our INT~WFC
survey of  M31, and  one of  our camera pointings  is centred  on this
field.  However,  the centre  of this field  is unsuitable  for direct
analysis  due to  severe crowding  and also  significant contamination
from M31.  Instead we use  the neighbouring INT WFC field, further out
on  the  M31  minor  axis,  which  still  contains  a  strong  NGC~205
population  but  with  much  less  crowding  and  a  much  weaker  M31
population.  Our reference field is the neighbouring field to this one,
again  further out  on  the minor  axis.   The resulting  CMD and  RGB
luminosity functions for NGC~205  are shown in Figure~\ref{ngc205} and
are constructed from objects lying  within $2\,-\,\sigma$ of the stellar
locus in  both filters.   The TRGB is  obvious in these  diagrams, and
application of our algorithm  yields the following results.  The group
of  stars  immediately brighter  than  the  tip,  visible in  the  RGB
luminosity function,  is most likely  a combination of bright  AGB and
M31  contamination.   Note  that  the metallicity  used  to  calculate
$M_I^{TRGB}$ for  this galaxy is  again not the median  metallicity of
the RGB but the representative metallicity of the strip of stars shown
in the left panel of Figure~\ref{ngc205} \mbox{([M/H] $\sim -1$).}

\begin{eqnarray*}
NGC~205:\\
{\rm I_{TRGB}}           & = & 20.65 \pm 0.03~{\rm mags} \nonumber\\
{\rm E\left(B - V\right)}& = & 0.062~{\rm mags}        \nonumber\\
{\rm \left[M/H\right]_{\alpha=0.0}}& = & -0.8               \nonumber\\
{\rm \left[M/H\right]_{\alpha=0.3}}&=&-0.7\nonumber\\
{\rm M_I^{TRGB}}         & = & -4.05~{\rm mags}     \nonumber\\
{\rm \left(m - M\right)_o} & = & 24.58 \pm 0.07~{\rm mags}\nonumber\\
{\rm D_{NGC~205}}        & = & 824 \pm 27~{\rm kpc}      \nonumber\\
\\
NGC~205~Error~Budget: \nonumber \\
{\rm Photometry}-~~~rms   & : & \pm 0.03~{\rm mags} \nonumber\\
{\rm ~~~~~~~~~~}~-zeropt   & : & \pm 0.02~{\rm mags} \nonumber\\
{\rm Reddening}    & : & \pm 0.02~{\rm mags} \nonumber\\
{\rm M_I^{TRGB}}   & : &  \pm 0.05~{\rm mags} \nonumber\\
{\rm Algorithm}    & : &  \pm 0.03~{\rm mags} \nonumber\\
{\rm Total}        & : &  \pm 0.07~{\rm mags} \nonumber\\
\end{eqnarray*}

\subsection{NGC~185}

NGC 185 ($0{\rm  h}~38{\rm m} ~58.0{\rm s}, +48^\circ  ~20' ~15''$) is
another  dwarf elliptical  companion to  M31.   \cite{lee1993b} found
strong evidence for three distinct stellar populations: a well defined
RGB, indicating a mean metallicity  of \mbox{${\rm [Fe/H]} = -1.23 \pm
0.16$,} with  a large  dispersion; a bright  AGB population  above the
TRGB, implying  the presence of a strong  intermediate age population;
and some young  stars with blue - yellow  colours, suggesting that NGC
185  has  Population~I  stars  as  well  as  Population~II  (see  also
\citealt{hodge1963,hodge1973,gallagher1981,johnson1983,wiklind1986}).

\begin{figure*}
\begin{center}
\includegraphics[angle=270, width=13.5cm]{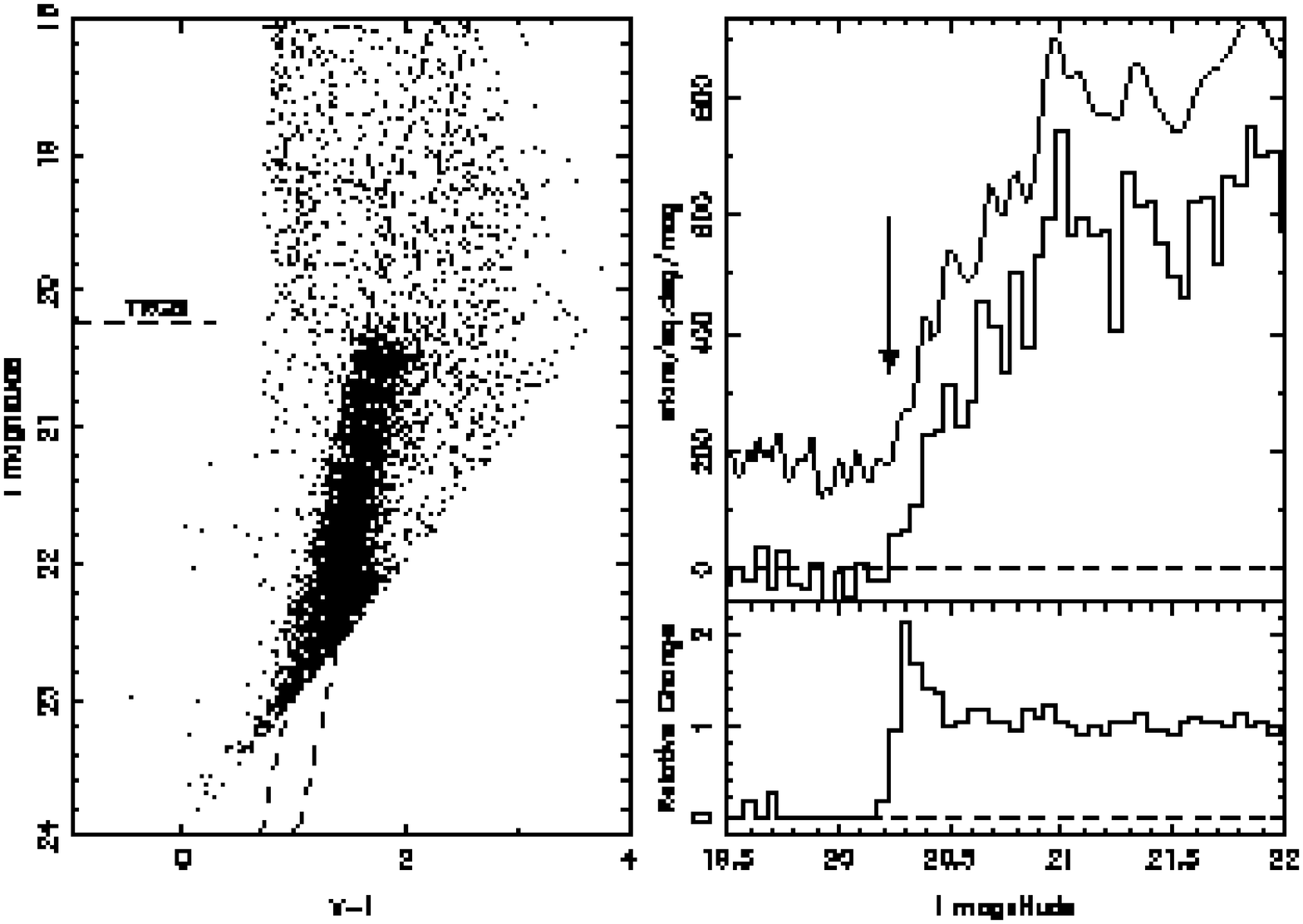}
\caption{NGC 185 CMD (left  panel), RGB luminosity function and offset
LPD (upper right panel), and heuristic signal (lower right panel). The
TRGB is measured to lie at ${\rm I} = 20.23$~mags and is marked on the
CMD by a horizontal dashed line  and on the luminosity functions by an
arrow.}
\label{ngc185}
\end{center}
\end{figure*}

The CMD and RGB  luminosity functions displayed in Figure~\ref{ngc185}
are constructed from  all the stars located further  than 0\degg1 from
the centre of NGC~185, but interior to 0\degg2. By only sampling stars
in a ring  around NGC~185, we negate the  possible effects of crowding
in the inner regions while still leaving a large enough area to act as
a reference field. The lack  of stars at magnitudes brighter than 20.2
in the resultant RGB luminosity function/LPD shows that the foreground
is well  removed by  this technique.  Additionally,  due to  the large
number of objects for which we have good photometry, we have used only
objects that  lie within  $2\,-\,\sigma$ of the  stellar locus  in both
filters.  Application of  our TRGB  algorithm to  the LPD  for NGC~185
gives the results detailed below, along with the error budget for this
galaxy.  The  value of  the reddening from  \cite{schlegel1998} agrees
well  with  that derived  independently  by  \cite{lee1993b} of  ${\rm
E\left(B - V\right)} = 0.19 \pm 0.03$~mags.

\begin{eqnarray*}
NGC~185:\\
{\rm I_{TRGB}}           & = & 20.23 \pm 0.03~{\rm mags} \nonumber\\
{\rm E\left(B - V\right)}& = & 0.179~{\rm mags}        \nonumber\\
{\rm \left[M/H\right]_{\alpha=0.0}}  & = & -1.2               \nonumber\\
{\rm \left[M/H\right]_{\alpha=0.3}}&=&-1.1\nonumber\\
{\rm M_I^{TRGB}}         & = & -4.065~{\rm mags}     \nonumber\\
{\rm \left(m - M\right)_o} & = & 23.95 \pm 0.09~{\rm mags}\nonumber\\
{\rm D_{NGC185}}        & = & 616 \pm 26~{\rm kpc}      \nonumber\\
\\
NGC~185~Error~Budget:\nonumber \\
{\rm Photometry}-~~~rms   & : & \pm 0.02~{\rm mags} \nonumber\\
{\rm ~~~~~~~~~~}~-zeropt   & : & \pm 0.02~{\rm mags} \nonumber\\
{\rm Reddening}    & : & \pm 0.06~{\rm mags} \nonumber\\
{\rm M_I^{TRGB}}   & : &  \pm 0.05~{\rm mags} \nonumber\\
{\rm Algorithm}    & : &  \pm 0.03~{\rm mags} \nonumber\\
{\rm Total}        & : &  \pm 0.09~{\rm mags} \nonumber\\
\end{eqnarray*}

\subsection{NGC 147}

NGC 147 ($0{\rm h} ~33{\rm  m} ~12.1{\rm s}, +48^\circ ~30' ~32''$) is
another dwarf  elliptical companions to  M31, and it is  recognised as
being the most unique in  terms of its stellar content. \cite{han1997}
conducted a thorough study of its stellar populations using the Hubble
Space Telescope's  (HST's) Wide  Field Planetary Camera  (WFPC2).  The
colour of the  RGB implied a metallicity of  ${\rm [Fe/H]}\simeq -0.9$
although radial variations were observed in the average RGB colour and
dispersion. Unlike  NGC~185 and NGC~205, no evidence  for Population I
stars  was  found,  although   an  intermediate  age  population  was
observed.

\begin{figure*}
\begin{center}
\includegraphics[angle=270, width=13.5cm]{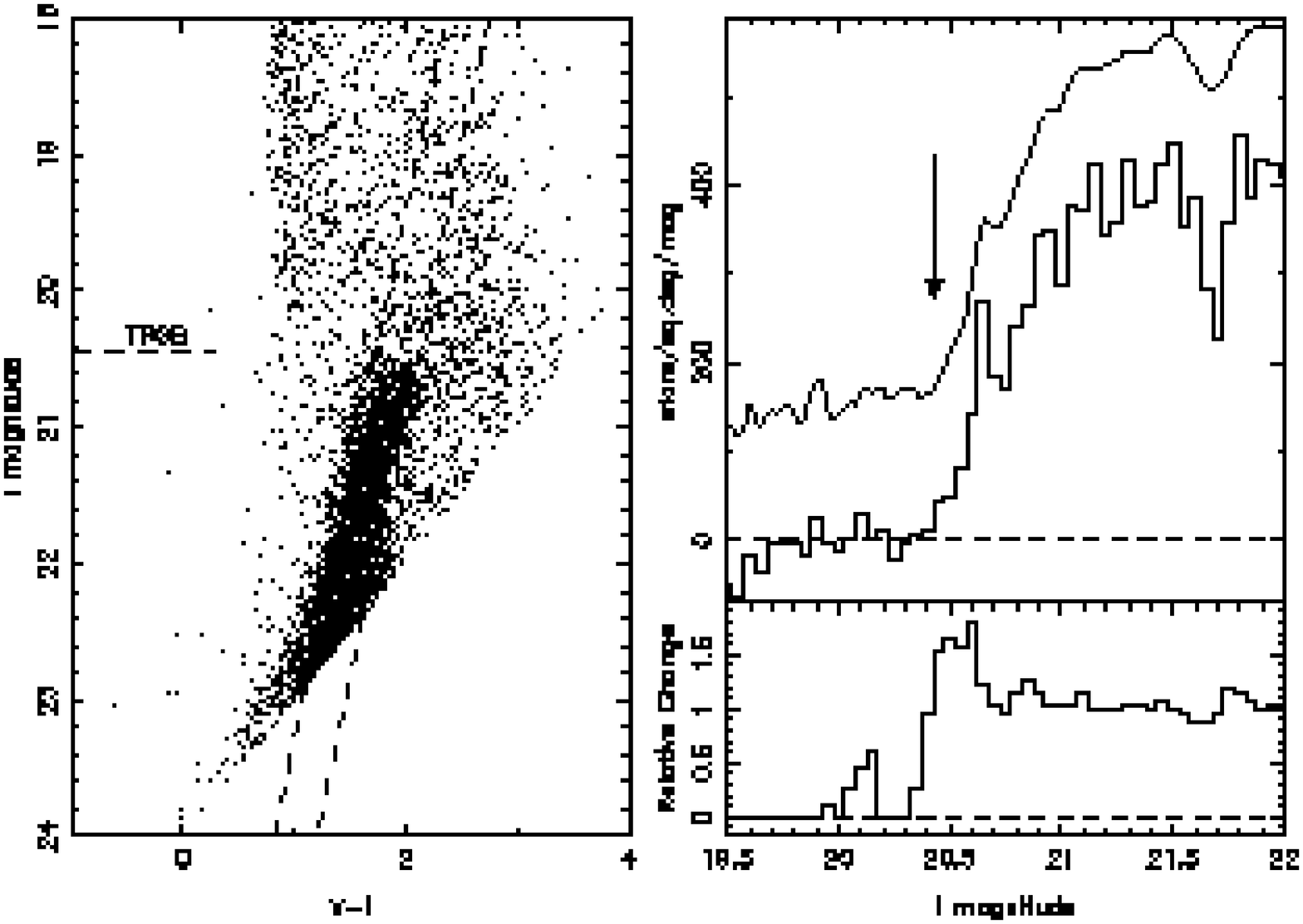}
\caption{NGC 147 CMD (left  panel), RGB luminosity function and offset
LPD (upper right panel), and heuristic signal (lower right panel). The
TRGB is measured to lie at ${\rm I} = 20.43$~mags and is marked on the
CMD by a horizontal dashed line  and on the luminosity functions by an
arrow.}
\label{ngc147}
\end{center}
\end{figure*}

The CMD and RGB  luminosity functions shown in Figure~\ref{ngc147} are
constructed from all  the stars located exterior to  0\degg15 from the
centre of NGC~147,  but within 0\degg25. In a  similar way to NGC~185,
this   helps  negate  the   possible  effects   of  crowding   on  our
measurement. Due to the number of objects with good photometry, we use
only   objects   lying   within   $2\,-\,\sigma$   of   the   stellar
locus. Inspection  of the RGB  luminosity function/LPD shows  that the
foreground  is minimal.  The result  of  the application  of our  TRGB
algorithm  to the  LPD  for NGC~147,  and  the error  budget for  this
galaxy, is shown below.

\begin{eqnarray*}
NGC~147\\
{\rm I_{TRGB}}           & = & 20.43 \pm 0.04~{\rm mags} \nonumber\\
{\rm E\left(B - V\right)}& = & 0.175~{\rm mags}        \nonumber\\
{\rm \left[M/H\right]_{\alpha=0.0}}  & = & -1.1               \nonumber\\
{\rm \left[M/H\right]_{\alpha=0.3}}&=&-1.0\nonumber\\
{\rm M_I^{TRGB}}         & = & -4.055~{\rm mags}     \nonumber\\
{\rm \left(m - M\right)_o} & = & 24.15 \pm 0.09~{\rm mags}\nonumber\\
{\rm D_{NGC147}}        & = & 675 \pm 27~{\rm kpc}      \nonumber\\
\\
NGC~147~Error~Budget&  & \nonumber \\
{\rm Photometry}-~~~rms   & : & \pm 0.02~{\rm mags} \nonumber\\
{\rm ~~~~~~~~~~}~-zeropt   & : & \pm 0.02~{\rm mags} \nonumber\\
{\rm Reddening}    & : & \pm 0.05~{\rm mags} \nonumber\\
{\rm M_I^{TRGB}}   & : &  \pm 0.05~{\rm mags} \nonumber\\
{\rm Algorithm}    & : &  \pm 0.04~{\rm mags} \nonumber\\
{\rm Total}        & : &  \pm 0.09~{\rm mags} \nonumber\\
\end{eqnarray*}

\subsection{Pegasus (DDO 216, UGC 12613)}

Pegasus ($23{\rm h} ~28{\rm m} ~36.2{\rm s}, +14^\circ ~44' ~35''$),
unlike the previous systems, is a small dwarf irregular galaxy,
discovered in the late 1950's by Wilson
(\citealt{holmberg1958}). \mbox{\cite{hoessel1982}} were the first to
study it in detail after the introduction of CCDs to astronomy; they
found a lack of any bright young stars but did infer the presence of
an old and intermediate stellar population from the presence of three
red star clusters. More recently, \cite{gallagher1998} have used WFPC2
in conjunction with ground-based data to study this galaxy. Their
results show an object that has a surprising mix of stellar
populations for such a small galaxy and suggest that Pegasus consists
predominantly of a young-to-intermediate age population.  This does
not affect our use of the TRGB as a distance indicator for this
galaxy, as the TRGB is of roughly constant magnitude for RGB stars
older than 2 Gyrs.

\begin{figure*}
\begin{center}
\includegraphics[angle=270, width=13.5cm]{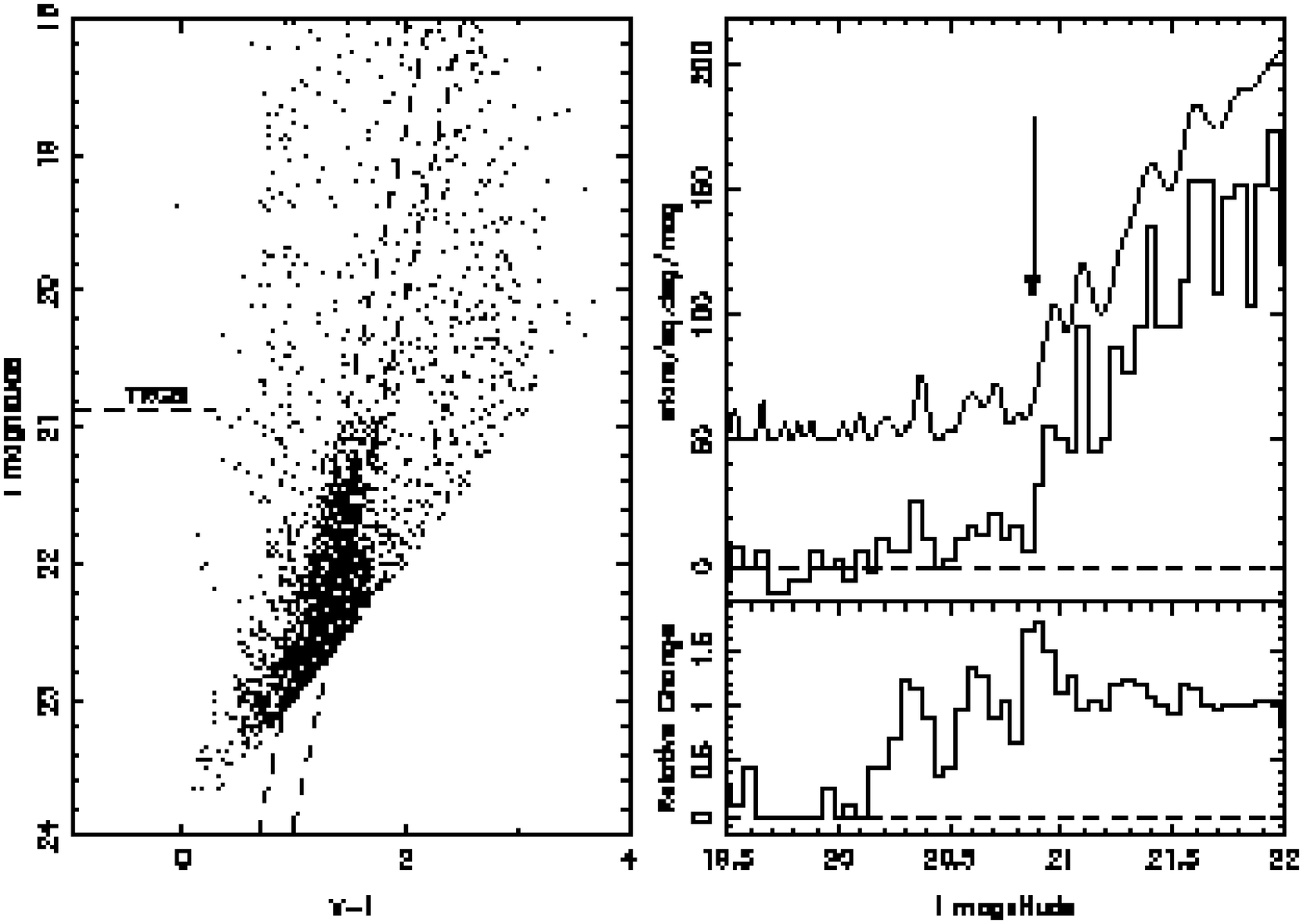}
\caption{Pegasus CMD (left panel),  RGB luminosity function and offset
LPD (upper right panel), and heuristic signal (lower right panel). The
TRGB is measured to lie at ${\rm I} = 20.87$~mags and is marked on the
CMD by a horizontal dashed line  and on the luminosity functions by an
arrow.}
\label{pegasus}
\end{center}
\end{figure*}

The  CMD  and  RGB  luminosity  functions for  Pegasus  are  shown  in
Figure~\ref{pegasus}.   These   were  constructed  by   using  objects
lying within  $3\,-\,\sigma$ of the stellar locus in  both filters,
located within 0\degg25 of the  centre of Pegasus.  To negate possible
crowding effects, we do not use  the inner 0\degg05.  The onset of the
RGB is evident at $\sim  20.9$~mags. The excess of stars brighter than
this, visible in the RGB  lumniosity function/LPD, is attributed to an
AGB population. The result of the application of our TRGB algorithm is
detailed below.

\begin{eqnarray*}
Pegasus\\
{\rm I_{TRGB}}           & = & 20.87 \pm 0.03~{\rm mags} \nonumber\\
{\rm E\left(B - V\right)}& = & 0.064~{\rm mags}        \nonumber\\
{\rm \left[M/H\right]_{\alpha=0.0}}  & = & -1.4              \nonumber\\
{\rm \left[M/H\right]_{\alpha=0.3}}&=&-1.2\nonumber\\
{\rm M_I^{TRGB}}         & = & -4.07~{\rm mags}     \nonumber\\
{\rm \left(m - M\right)_o} & = & 24.82 \pm 0.07~{\rm mags}\nonumber\\
{\rm D_{Pegasus}}        & = & 919 \pm 30~{\rm kpc}      \nonumber\\
\\
Pegasus~Error~Budget&  & \nonumber \\
{\rm Photometry}-~~~rms   & : & \pm 0.03~{\rm mags} \nonumber\\
{\rm ~~~~~~~~~~}~-zeropt   & : & \pm 0.02~{\rm mags} \nonumber\\
{\rm Reddening}    & : & \pm 0.02~{\rm mags} \nonumber\\
{\rm M_I^{TRGB}}   & : &  \pm 0.05~{\rm mags} \nonumber\\
{\rm Algorithm}    & : &  \pm 0.03~{\rm mags} \nonumber\\
{\rm Total}        & : &  \pm 0.07~{\rm mags} \nonumber\\
\end{eqnarray*}

\subsection{WLM (DDO 221)}

WLM (Wolf-Lundmark-Melotte), located at  $0{\rm h} ~1{\rm m} ~58.1{\rm
s}, -15^\circ ~27' ~39''$, is  a dwarf irregular galaxy, discovered at
the start of the $20^{{\rm th}}$ century by \cite{wolf1910}. In recent
decades,  several groups  have analysed  the stellar  content  of this
object     (\citealt{sandage1985,ferraro1989,minniti1997,rejkuba2000}).
\cite{minniti1997} derive a mean  metallicity of ${\rm [Fe/H]} = -1.45
\pm  0.2$  from  the colour  of  the  RGB,  and  show that  young  and
intermediate  age stellar  populations are  present. The  discovery of
blue  horizontal  branch  (HB)   stars  by  \cite{rejkuba2000}  is  an
unambiguous indicator  of an  old stellar population.   H\,II regions,
Cepheids   and    planetary   nebulae   have    also   been   detected
(\citealt{skillman1989,sandage1985,jacoby1981}) and it would seem that
WLM consists  of a  disk-like component with possibly an old  extended 
stellar halo.

\begin{figure*}
\begin{center}
\includegraphics[angle=270, width=13.5cm]{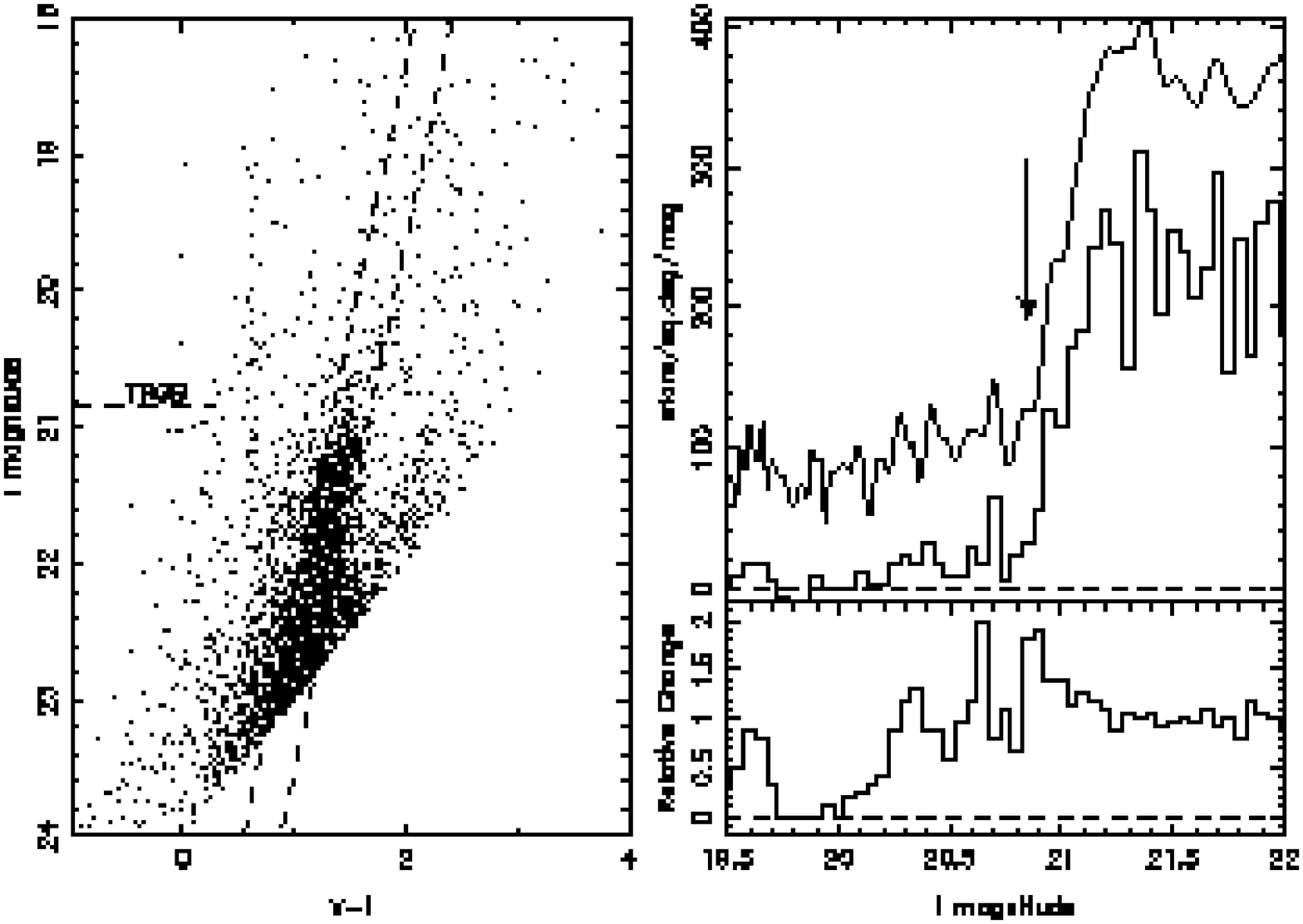}
\caption{WLM CMD (left panel),  RGB luminosity function and offset LPD
(upper  right panel), and  heuristic signal  (lower right  panel). The
TRGB is measured to lie at ${\rm I} = 20.85$~mags and is marked on the
CMD by a horizontal dashed line  and on the luminosity functions by an
arrow.}
\label{wlm}
\end{center}
\end{figure*}

Figure~\ref{wlm}  shows  the  CMD  and RGB  luminosity  functions  for
WLM. These were constructed from all stars located within 0\degg2 from
the centre of this galaxy.  Again, stars within 0\degg05 of the centre
were not used to minimise  potential problems due to crowding. The RGB
is clearly visible  in the CMD and an extended AGB  also appears to be
present.   This  is  most   clearly  visible  in  the  RGB  luminosity
function/LPD. The location of the tip, as derived by our algorithm, is
detailed below. The  slightly larger error on this  measurement is due
to the TRGB not being as  clearly defined as in the previous examples,
as inspection of the CMD shows.

\begin{eqnarray*}
WLM\\
{\rm I_{TRGB}}           & = & 20.85 \pm 0.05~{\rm mags} \nonumber\\
{\rm E\left(B - V\right)}& = & 0.035~{\rm mags}        \nonumber\\
{\rm \left[M/H\right]_{\alpha=0.0}}  & = & -1.5      \nonumber\\
{\rm \left[M/H\right]_{\alpha=0.3}}&=&-1.4\nonumber\\
{\rm M_I^{TRGB}}         & = & -4.065~{\rm mags}     \nonumber\\
{\rm \left(m - M\right)_o} & = & 24.85 \pm 0.08~{\rm mags}\nonumber\\
{\rm D_{WLM}}        & = & 932 \pm 33~{\rm kpc}      \nonumber\\
\\
WLM~Error~Budget   &  & \nonumber \\
{\rm Photometry}-~~~rms   & : & \pm 0.02~{\rm mags} \nonumber\\
{\rm ~~~~~~~~~~}~-zeropt   & : & \pm 0.02~{\rm mags} \nonumber\\
{\rm Reddening}    & : & \pm 0.01~{\rm mags} \nonumber\\
{\rm M_I^{TRGB}}   & : &  \pm 0.05~{\rm mags} \nonumber\\
{\rm Algorithm}    & : &  \pm 0.05~{\rm mags} \nonumber\\
{\rm Total}        & : &  \pm 0.08~{\rm mags} \nonumber\\
\end{eqnarray*}

\subsection{LGS3 (Pisces Dwarf)}

LGS3 ($1{\rm h} ~3{\rm m} ~52.9{\rm s}, +21^\circ ~53' ~05''$) is a
very faint member of the Local Group, originally discovered by
\cite{karachentseva1976} (see also \citealt{kowal1978,thuan1979}).
\cite{lee1995} conducted a study of the stellar populations; an RGB is
evident and its colour implies a metallicity of ${\rm [Fe/H]} = -2.10
\pm 0.22$. A small number of bright AGB stars and several blue stars
are also observed.  Along with the Phoenix dwarf galaxy, LGS3 is
suspected of being a transition-type galaxy, somewhere between a dwarf
irregular and a dwarf spheroidal galaxy
(\citealt{cook1989,vanderydt1991}). It could be that this is a real
transition object, but the possibility that it is merely a rare, but
statistically acceptable, version of one of the two classes must also
be considered (see \citealt{aparicio1997} and \citealt{lee1995} for
differing conclusions on this matter).

\begin{figure*}
\begin{center}
\includegraphics[angle=270, width=13.5cm]{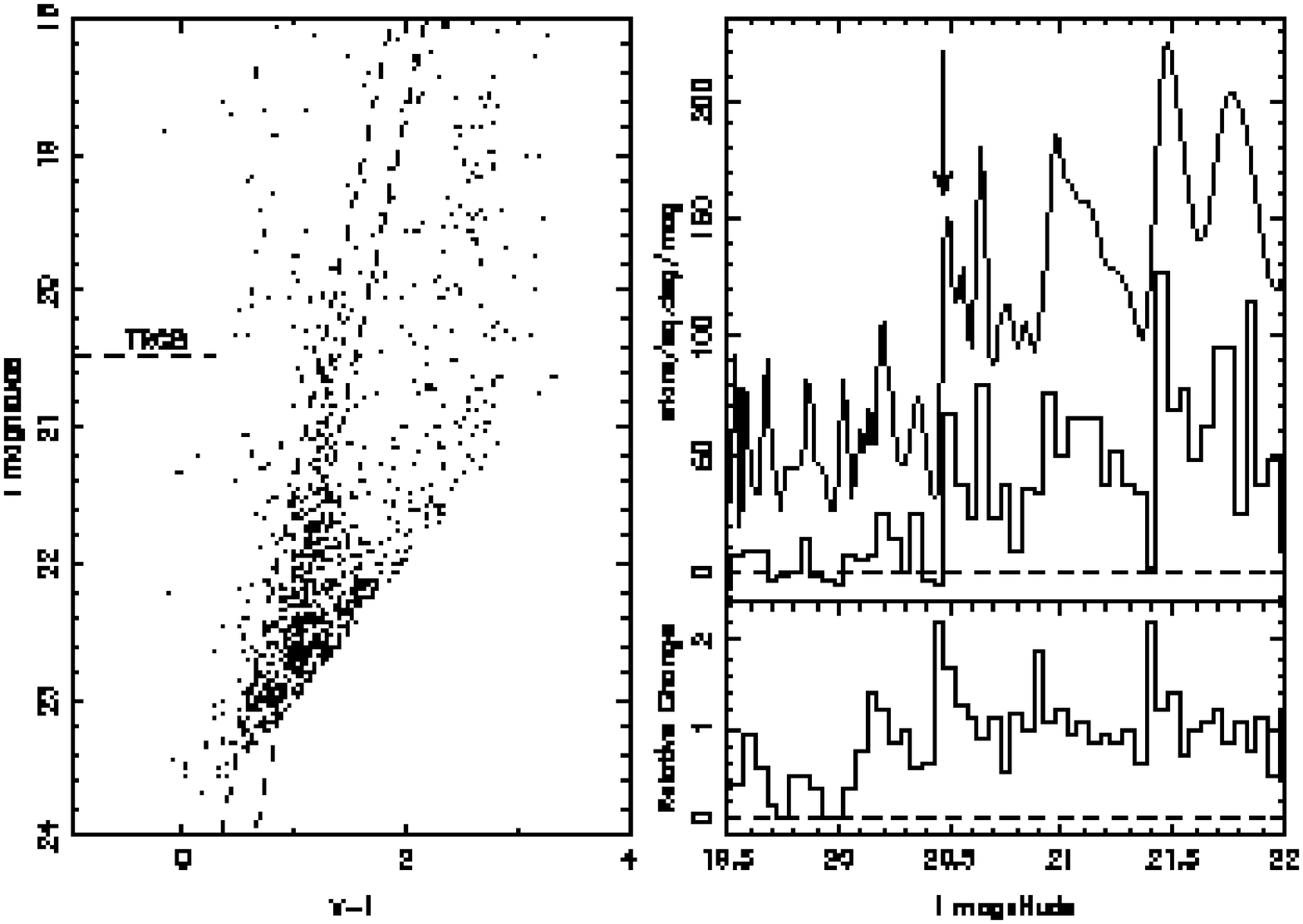}
\caption{LGS3 CMD (left panel), RGB luminosity function and offset LPD
(upper  right panel), and  heuristic signal  (lower right  panel). The
TRGB is measured to lie at ${\rm I} = 20.47$~mags and is marked on the
CMD by a horizontal dashed line  and on the luminosity functions by an
arrow.}
\label{lgs3}
\end{center}
\end{figure*}

The  CMD  and   RGB  luminosity  functions  for  LGS3   are  shown  in
Figure~\ref{lgs3}. They have been constructed from all stellar objects
located within 0\degg15 of the centre  of this galaxy.  LGS3 is a much
fainter galaxy  than the  previous ones  and so its  CMD is  much more
sparsely  populated.   As such,  its  RGB  luminosity function/LPD  is
relatively noisy. A  large increase in the number  of stars is readily
observable,  however, at  $\sim 20.5$~mag  which we  attribute  to the
TRGB. Below are the results  from the application of our algorithm. We
have been  unable to calculate the  full MDF for LGS3  because a large
fraction of  the stars in  LGS3 appear at  least as metal poor  as our
most metal poor  evolutionary tracks.  We therefore adopt  this as the
typical  metallicity  of   the  system.  

\begin{eqnarray*}
LGS3\\
{\rm I_{TRGB}}           & = & 20.47 \pm 0.03~{\rm mags} \nonumber\\
{\rm E\left(B - V\right)}& = & 0.042~{\rm mags}        \nonumber\\
{\rm \left[M/H\right]_{\alpha=0.0}}  & \lesssim& -2.3           \nonumber\\
{\rm \left[M/H\right]_{\alpha=0.3}}&\lesssim&-2.0\nonumber\\
{\rm M_I^{TRGB}}         & = & -4.04~{\rm mags}     \nonumber\\
{\rm \left(m - M\right)_o} & = & 24.43 \pm 0.07~{\rm mags}\nonumber\\
{\rm D_{LGS3}}        & = & 769 \pm 23~{\rm kpc}      \nonumber\\
\\
LGS3~Error~Budget  &  & \nonumber \\
{\rm Photometry}-~~~rms   & : & \pm 0.02~{\rm mags} \nonumber\\
{\rm ~~~~~~~~~~}~-zeropt   & : & \pm 0.02~{\rm mags} \nonumber\\
{\rm Reddening}    & : & \pm 0.01~{\rm mags} \nonumber\\
{\rm M_I^{TRGB}}   & : &  \pm 0.05~{\rm mags} \nonumber\\
{\rm Algorithm}    & : &  \pm 0.03~{\rm mags} \nonumber\\
{\rm Total}        & : &  \pm 0.07~{\rm mags} \nonumber\\
\end{eqnarray*}

\subsection{Cetus}

Cetus ($0{\rm h} ~26{\rm m} ~11{\rm s}, -11^\circ ~2' ~40''$) was only
recently discovered  by \cite{whiting1999} as part of  the same survey
that  also led  to the  discovery of  a Local  Group galaxy  in Antlia
(\citealt{whiting1997}).    \cite{sarajedini2002}  have   conducted  a
survey of the stellar populations of this dwarf spheroidal using HST's
WFPC2  instrument. The  colour of  the RGB  suggests a  metallicity of
${\rm [Fe/H]} = -1.7$, with a dispersion of $\simeq 0.2$. A strong red
HB is  observed with indications  of a less numerous  bluer component.
The HB is observed to  be significantly redder than those for globular
clusters  of similar  metallicities.   They conclude  that  this is  a
manifestation of  the second-parameter effect,  and if it  is accepted
that this is due  to age, then Cetus is 2 -  3 Gyrs younger than these
globular clusters.

\begin{figure*}
\begin{center}
\includegraphics[angle=270, width=13.5cm]{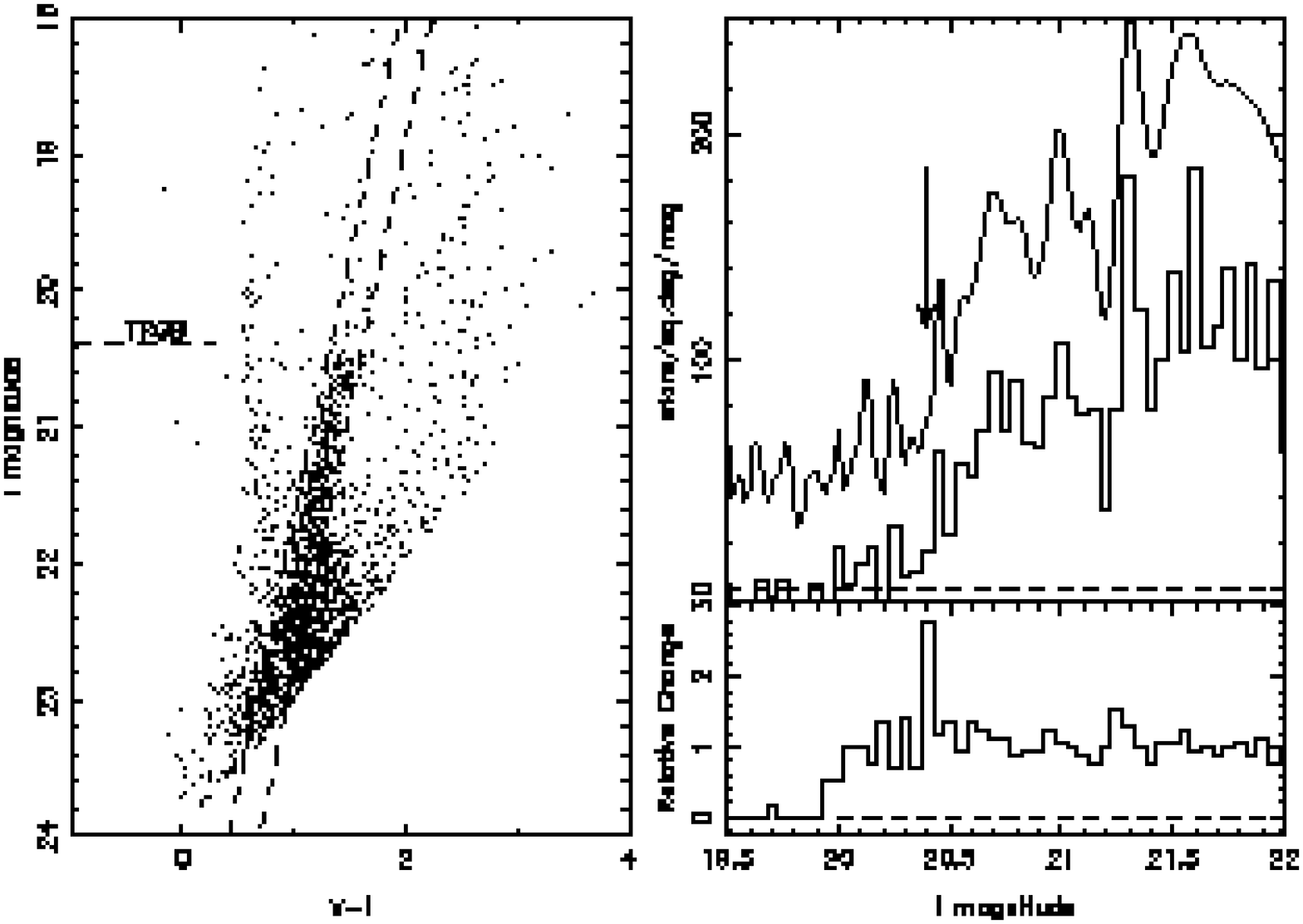}
\caption{Cetus CMD  (left panel),  RGB luminosity function  and offset
LPD (upper right panel), and heuristic signal (lower right panel). The
TRGB is measured to lie at ${\rm I} = 20.39$~mags and is marked on the
CMD by a horizontal dashed line  and on the luminosity functions by an
arrow.}
\label{cetus}
\end{center}
\end{figure*}

The  CMD   and  RGB  luminosity   function  for  Cetus  is   shown  in
Figure~\ref{cetus}. This has been constructed from all stellar objects
located within 0\degg2  from the centre of this  galaxy, and the onset
of the  RGB can be clearly  seen. Application of our  algorithm to the
LPD yields the results presented below.

\begin{eqnarray*}
Cetus\\
{\rm I_{TRGB}}           & = & 20.39 \pm 0.03~{\rm mags} \nonumber\\
{\rm E\left(B - V\right)}& = & 0.029~{\rm mags}        \nonumber\\
{\rm \left[M/H\right]_{\alpha=0.0}}  & = & -1.6              \nonumber\\
{\rm \left[M/H\right]_{\alpha=0.3}}&=&-1.5\nonumber\\
{\rm M_I^{TRGB}}         & = & -4.055~{\rm mags}     \nonumber\\
{\rm \left(m - M\right)_o}& = & 24.39 \pm 0.07~{\rm mags}\nonumber\\
{\rm D_{Cetus}}        & = & 755 \pm 23~{\rm kpc}      \nonumber\\
\\
Cetus~Error~Budget&  &\nonumber \\
{\rm Photometry}-~~~rms   & : & \pm 0.02~{\rm mags} \nonumber\\
{\rm ~~~~~~~~~~}~-zeropt   & : & \pm 0.02~{\rm mags} \nonumber\\
{\rm Reddening}    & : & \pm 0.01~{\rm mags} \nonumber\\
{\rm M_I^{TRGB}}   & : &  \pm 0.05~{\rm mags} \nonumber\\
{\rm Algorithm}    & : &  \pm 0.03~{\rm mags} \nonumber\\
{\rm Total}        & : &  \pm 0.07~{\rm mags} \nonumber\\
\end{eqnarray*}

\subsection{Andromeda III}

The  Andromeda~III  dwarf  spheroidal  galaxy ($0{\rm  h}  ~35{\rm  m}
~33.8{\rm   s},    +36^\circ   ~29'   ~52''$)    was   discovered   by
\citealt{vandenbergh1972a} (see also \citealt{vandenbergh1972b}) along
with  Andromeda~I   and  II,   and  the  background   dwarf  irregular
Andromeda~IV.   \cite{dacosta2002}  have  used  WFPC2 to  obtain  deep
photometry as part  of their HST survey of  the stellar populations of
M31's dwarf spheroidal satellites (\citealt{dacosta1996,dacosta2000}).
By comparing the RGB with those of standard globular clusters, And~III
is found to have a mean metallicity of ${\rm [Fe/H]} \simeq -1.88$ and
a dispersion  of 0.12 (see  also \citealt{armandroff1993}).  It  has a
very obvious  HB, which is in fact  redder than that in  both And~I \&
II. Since it  is also metal-poorer than both  these systems, then this
is interpreted as being due  to age effects.  If correct, And~III must
therefore be younger than the globular clusters by $\sim 3$ Gyrs.  The
interested reader is referred to Figure~10 in \cite{dacosta2002} which
has an excellent comparison of their deep CMDs for And~I, II \& III.

\begin{figure*}
\begin{center}
\includegraphics[angle=270, width=13.5cm]{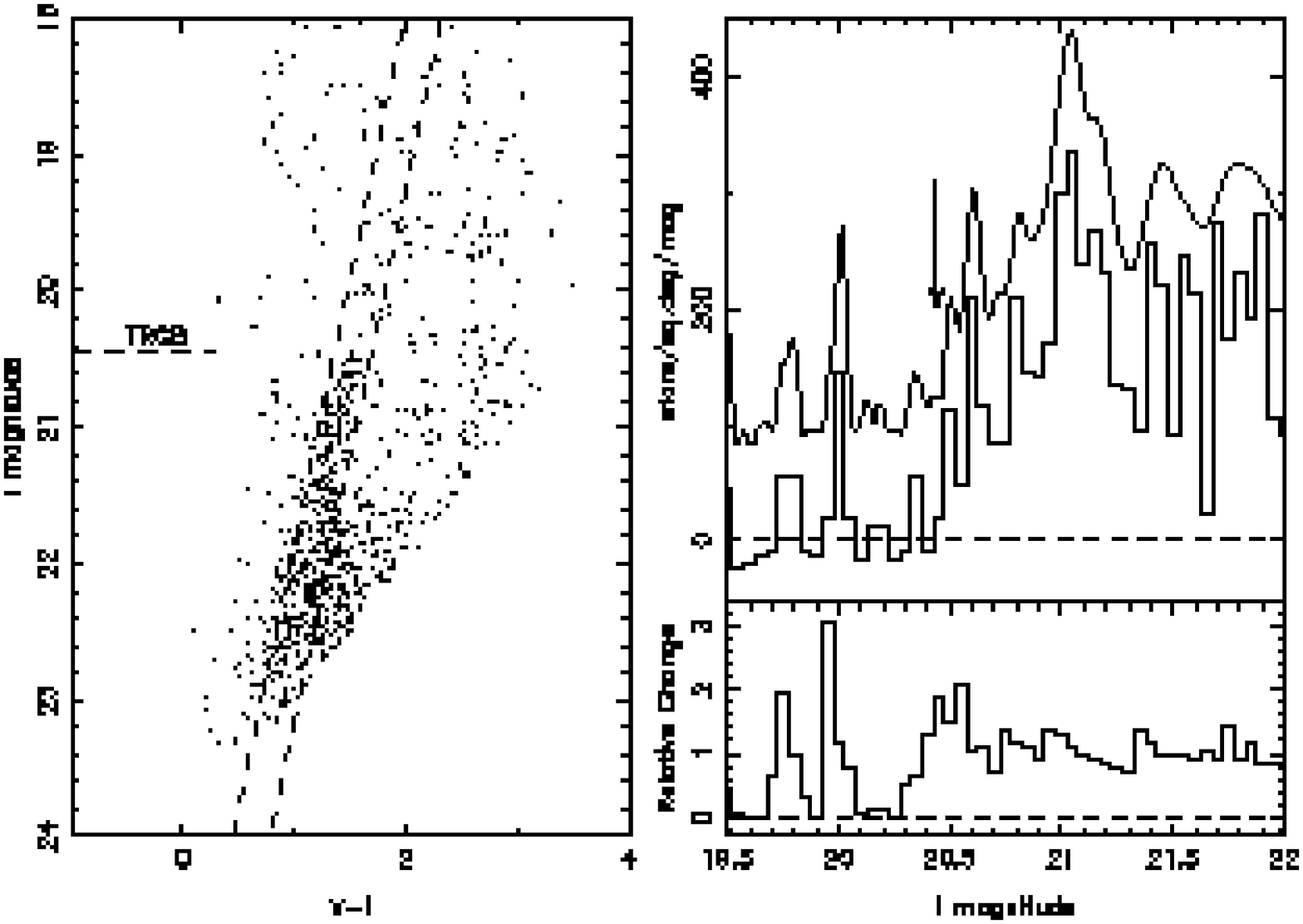}
\caption{And III CMD (left  panel), RGB luminosity function and offset
LPD (upper right panel), and heuristic signal (lower right panel). The
TRGB is measured to lie at ${\rm I} = 20.44$~mags and is marked on the
CMD by a horizontal dashed line  and on the luminosity functions by an
arrow.}
\label{andiii}
\end{center}
\end{figure*}

Figure~\ref{andiii} shows the CMD  and RGB luminosity function for all
stellar objects  within 0\degg1 of the  centre of And  III.  A smaller
area is used so  as to more clearly detect the RGB  over and above the
foreground stars.   The CMD and RGB luminosity  functions are slightly
noisy due  to the faintness of  the galaxy, although the  onset of the
TRGB     is    readily     visible    in     the     RGB    luminosity
function/LPD.  Application of  our TRGB  algorithm yields  the results
detailed below.

\begin{eqnarray*}
And~III\\
{\rm I_{TRGB}}           & = & 20.44 \pm 0.04~{\rm mags} \nonumber\\
{\rm E\left(B - V\right)}& = & 0.058~{\rm mags}        \nonumber\\
{\rm \left[M/H\right]_{\alpha=0.0}}  & = & -1.7       \nonumber\\
{\rm \left[M/H\right]_{\alpha=0.3}}&=&-1.6\nonumber\\
{\rm M_I^{TRGB}}         & = & -4.045~{\rm mags}     \nonumber\\
{\rm \left(m - M\right)_o} & = & 24.37 \pm 0.07~{\rm mags}\nonumber\\
{\rm D_{And~III}}        & = & 749 \pm 24~{\rm kpc}      \nonumber\\
\\
And~III~Error~Budget&  & \nonumber \\
{\rm Photometry}-~~~rms   & : & \pm 0.02~{\rm mags} \nonumber\\
{\rm ~~~~~~~~~~}~-zeropt   & : & \pm 0.02~{\rm mags} \nonumber\\
{\rm Reddening}    & : & \pm 0.02~{\rm mags} \nonumber\\
{\rm M_I^{TRGB}}   & : &  \pm 0.05~{\rm mags} \nonumber\\
{\rm Algorithm}    & : &  \pm 0.04~{\rm mags} \nonumber\\
{\rm Total}        & : &  \pm 0.07~{\rm mags} \nonumber\\
\end{eqnarray*}

\subsection{Andromeda V, Andromeda VI (Pegasus dSph) and Andromeda VII (Cassiopeia dSph)}

Andromeda  V  ($1{\rm  h}  ~10{\rm  m} ~17.1{\rm  s},  +47^\circ  ~37'
~41''$),  VI  ($23{\rm h}  ~51{\rm  m}  ~46.3{\rm  s}, +24^\circ  ~34'
~57''$),  \& VII  ($23{\rm h}  ~26{\rm m}  ~31{\rm s},  +50^\circ ~41'
~31''$)  were, prior  to the  discovery  of And~IX,  the three  newest
additions  to the  M31  subgroup.   And~V and  VI  were discovered  by
\cite{armandroff1998,armandroff1999}  after   they  had  performed  an
analysis on  the disgitised version  of the second Palomar  Sky Survey
(POSS-II; \citealt{reid1991,reid1993,lasker1993}).   At the same time,
\cite{karachentsev1999}  independently  discovered  And~V (which  they
named  the  Pegasus  dSph)  as  well as  yet  another  companion,  the
Cassiopeia  dSph (also  named And  ~VII)  from their  analysis of  the
POSS-II  plates. All of  these satellites  appear to  resemble typical
dwarf  spheroidals,   with  little  or   no  evidence  for   young  or
intermediate                    stellar                    populations
(\citealt{armandroff1998,armandroff1999,grebel1999a,hopp1999,davidge2002}). Their
metallicities, as  derived from  the colour of  their RGBs,  are ${\rm
[Fe/H]}  \simeq -1.3$  for  And VI  (\citealt{grebel1999a}), and  ${\rm
[Fe/H]} \simeq  -1.4$ for  And VII. And  V was originally  measured to
have   a    metallicity   of    ${\rm   [Fe/H]}   \simeq    -1.5$   by
\cite{armandroff1998},    but   a    more   recent    measurement   by
\cite{davidge2002}  measures this object  as being  significantly more
metal  poor,  at  ${\rm [Fe/H]}  =  -2.2  \pm  0.1$. The  more  recent
measurement places this object firmly on the expected relation between
integrated  luminosity and metallicity  for dwarf  spheroidal galaxies
(eg. see Figure~4 of \citealt{caldwell1999}).

\begin{figure*}
\begin{center}
\includegraphics[angle=270, width=13.5cm]{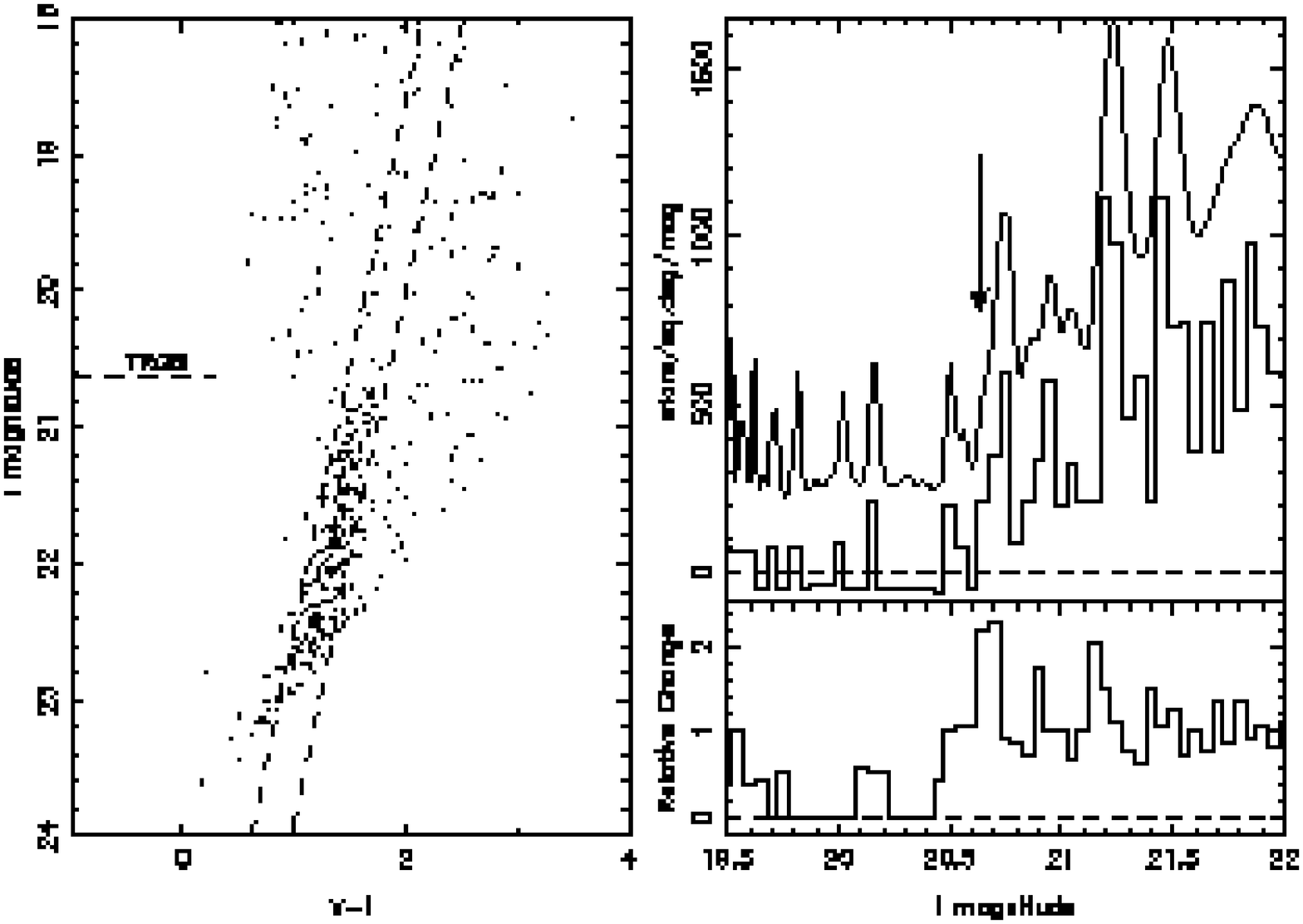}
\caption{And V  CMD (left panel),  RGB luminosity function  and offset
LPD (upper right panel), and heuristic signal (lower right panel). The
TRGB is measured to lie at ${\rm I} = 20.63$~mags and is marked on the
CMD by a horizontal dashed line  and on the luminosity functions by an
arrow.}
\label{andv}
\end{center}
\end{figure*}

Stars within $0\degg05$  ($\simeq 2.5 r_c$; \citealt{caldwell1999}) of
the centre of  And~V are used to construct the  CMD and RGB luminosity
functions shown  in Figure ~\ref{andv}. The result  of the application
of our  algorithm to  the LPD  is detailed below.

\begin{eqnarray*}
And~V\\
{\rm I_{TRGB}}           & = & 20.63 \pm 0.04~{\rm mags} \nonumber\\
{\rm E\left(B - V\right)}& = & 0.124~{\rm mags}          \nonumber\\
{\rm \left[M/H\right]_{\alpha=0.0}}  & = & -1.6            \nonumber\\
{\rm \left[M/H\right]_{\alpha=0.3}}&=&-1.5\nonumber\\
{\rm M_I^{TRGB}}         & = & -4.055~{\rm mags}          \nonumber\\
{\rm \left(m - M\right)_o} & = & 24.44 \pm 0.08~{\rm mags}\nonumber\\
{\rm D_{And~V}}          & = & 774 \pm 28~{\rm kpc}      \nonumber\\
\\
And~V~Error~Budget&  & \nonumber \\
{\rm Photometry}-~~~rms   & : & \pm 0.02~{\rm mags} \nonumber\\
{\rm ~~~~~~~~~~}~-zeropt   & : & \pm 0.02~{\rm mags} \nonumber\\
{\rm Reddening}    & : & \pm 0.04~{\rm mags} \nonumber\\
{\rm M_I^{TRGB}}   & : & \pm 0.05~{\rm mags} \nonumber\\
{\rm Algorithm}    & : & \pm 0.04~{\rm mags} \nonumber\\
{\rm Total}        & : & \pm 0.08~{\rm mags} \nonumber\\
\end{eqnarray*}

\begin{figure*}
\begin{center}
\includegraphics[angle=270, width=13.5cm]{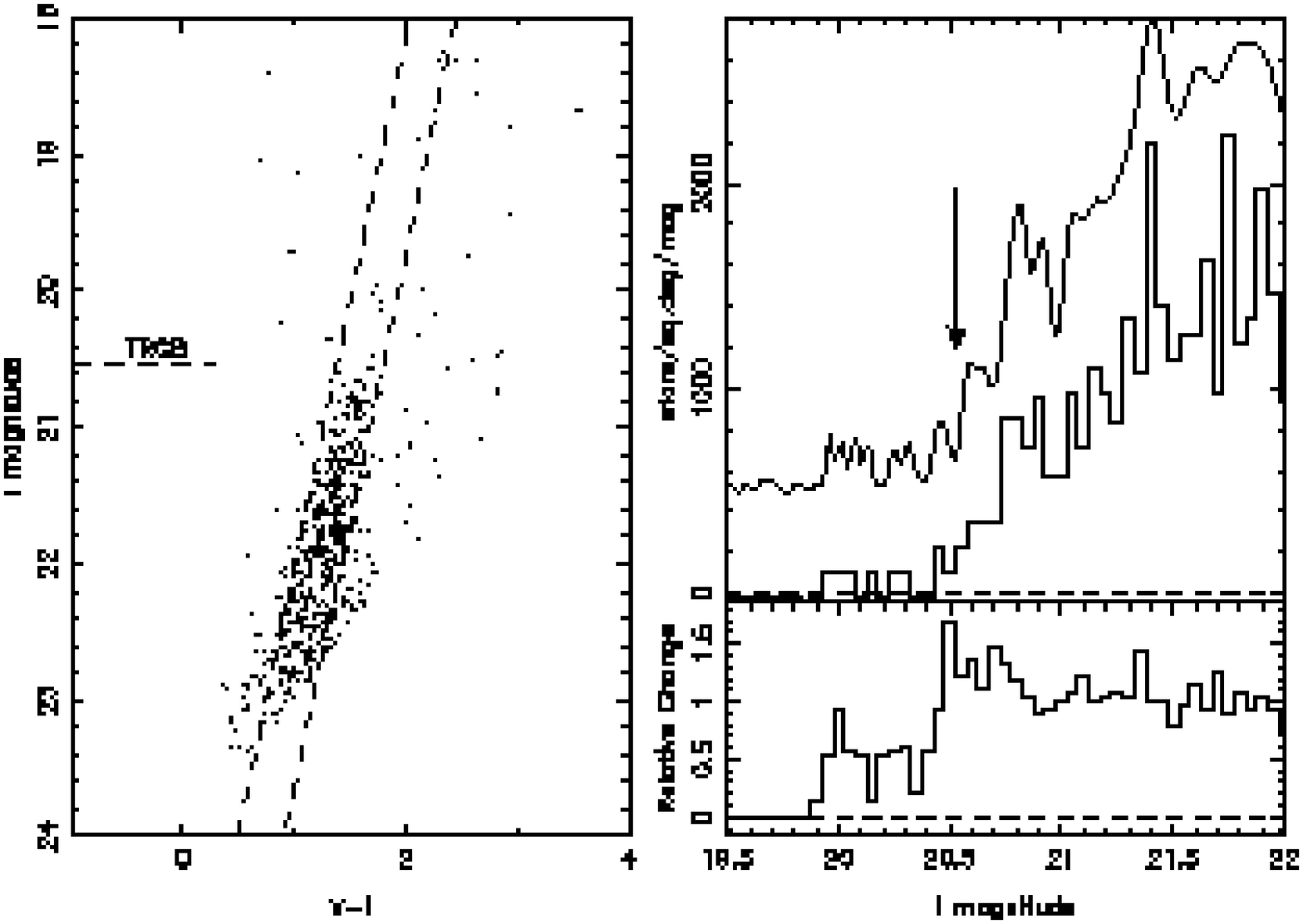}
\caption{And VI  CMD (left panel), RGB luminosity  function and offset
LPD (upper right panel), and heuristic signal (lower right panel). The
TRGB is measured to lie at ${\rm I} = 20.53$~mags and is marked on the
CMD by a horizontal dashed line  and on the luminosity functions by an
arrow.}
\label{andvi}
\end{center}
\end{figure*}

The CMD and  RGB luminosity function for all  the stars located within
0\degg05 from the centre of And~VI is shown in Figure~\ref{andvi}. The
result  of the application  of our  algorithm to  the LPD  is detailed
below; the few stars  in the luminosity functions immediately brighter
than the tip can be seen to be unassociated with the RGB by inspection
of the CMD. It may be that these stars are foreground contamination or
some extended AGB component.

\begin{eqnarray*}
And~VI\\
{\rm I_{TRGB}}           & = & 20.53 \pm 0.04~{\rm mags} \nonumber\\
{\rm E\left(B - V\right)}& = & 0.065~{\rm mags}        \nonumber\\
{\rm \left[M/H\right]_{\alpha=0.0}}  & = & -1.5             \nonumber\\
{\rm \left[M/H\right]_{\alpha=0.3}}&=&-1.3\nonumber\\
{\rm M_I^{TRGB}}         & = & -4.065~{\rm mags}     \nonumber\\
{\rm \left(m - M\right)_o} & = & 24.47 \pm 0.07~{\rm mags}\nonumber\\
{\rm D_{And~VI}}        & = & 783 \pm 25~{\rm kpc}      \nonumber\\
\\
And~VI~Error~Budget&  &\nonumber \\
{\rm Photometry}-~~~rms   & : & \pm 0.02~{\rm mags} \nonumber\\
{\rm ~~~~~~~~~~}~-zeropt   & : & \pm 0.02~{\rm mags} \nonumber\\
{\rm Reddening}    & : & \pm 0.02~{\rm mags} \nonumber\\
{\rm M_I^{TRGB}}   & : &  \pm 0.05~{\rm mags} \nonumber\\
{\rm Algorithm}    & : &  \pm 0.04~{\rm mags} \nonumber\\
{\rm Total}        & : &  \pm 0.07~{\rm mags} \nonumber\\
\end{eqnarray*}

\begin{figure*}
\begin{center}
\includegraphics[angle=270, width=13.5cm]{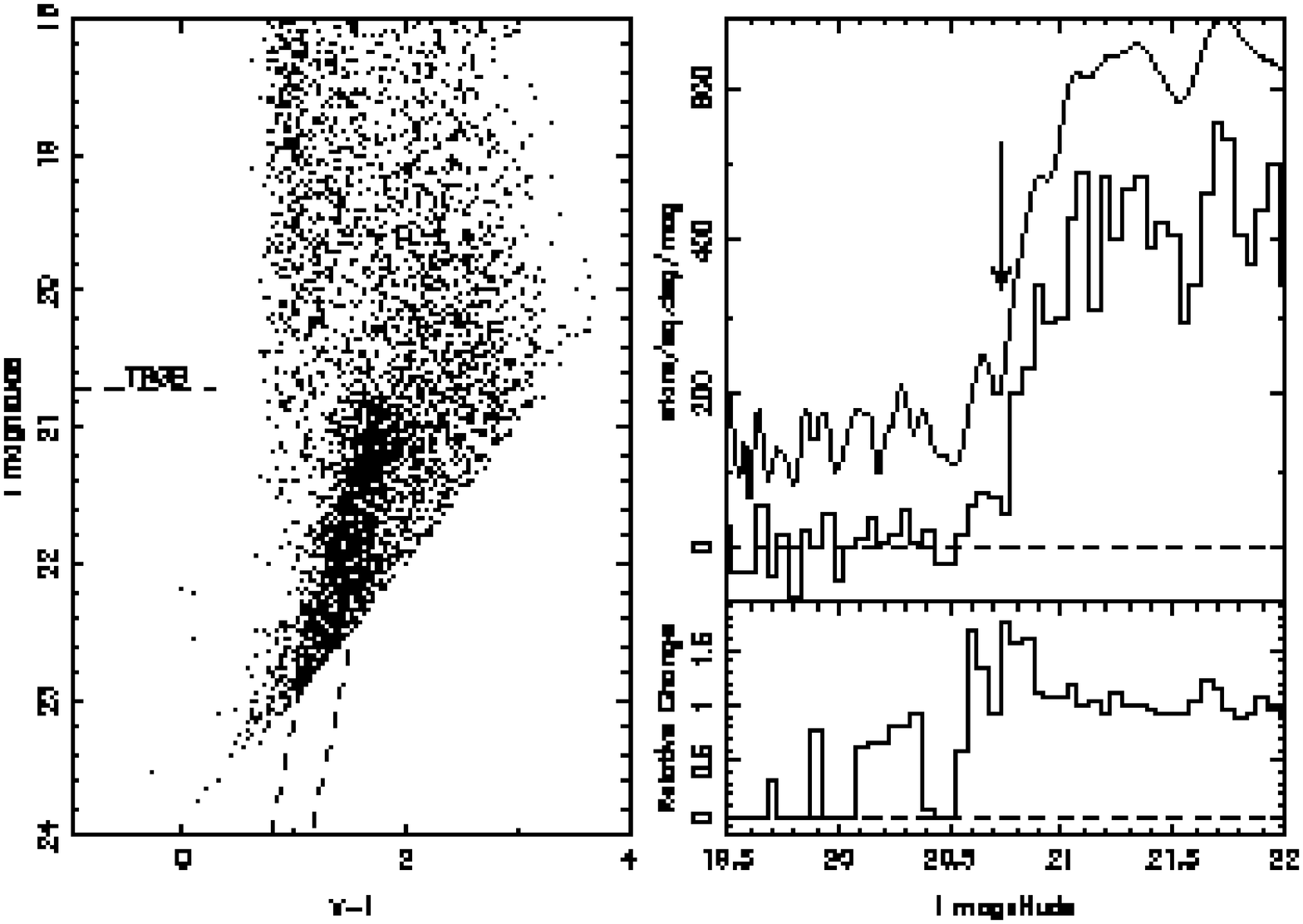}
\caption{And VII CMD (left  panel), RGB luminosity function and offset
LPD (upper right panel), and heuristic signal (lower right panel). The
TRGB is measured to lie at ${\rm I} = 20.73$~mags and is marked on the
CMD by a horizontal dashed line  and on the luminosity functions by an
arrow.}
\label{andvii}
\end{center}
\end{figure*}

Figure~\ref{andvii}  shows the  CMD  and RGB  luminosity function  for
And~VII. These are constructed from all stars within 0\degg15 from the
centre  of this  galaxy. The  result of  the TRGB  algorithm  is shown
below. The grouping  of stars brighter than the  proposed tip location
are most  likely an AGB population;  inspection of the  CMD shows that
this grouping is unlikely to be associated with the TRGB.

\begin{eqnarray*}
And~VII\\
{\rm I_{TRGB}}           & = & 20.73 \pm 0.05~{\rm mags} \nonumber\\
{\rm E\left(B - V\right)}& = & 0.199~{\rm mags}        \nonumber\\
{\rm \left[M/H\right]_{\alpha=0.0}}  & = & -1.4             \nonumber\\
{\rm \left[M/H\right]_{\alpha=0.3}}&=&-1.3\nonumber\\
{\rm M_I^{TRGB}}         & = & -4.07~{\rm mags}     \nonumber\\
{\rm \left(m - M\right)_o} & = & 24.41 \pm 0.10~{\rm mags}\nonumber\\
{\rm D_{And~VII}}        & = & 763 \pm 35~{\rm kpc}      \nonumber\\
\\
And~VII~Error~Budget&  & \nonumber \\
{\rm Photometry}-~~~rms   & : & \pm 0.03~{\rm mags} \nonumber\\
{\rm ~~~~~~~~~~}~-zeropt   & : & \pm 0.02~{\rm mags} \nonumber\\
{\rm Reddening}    & : & \pm 0.06~{\rm mags} \nonumber\\
{\rm M_I^{TRGB}}   & : &  \pm 0.05~{\rm mags} \nonumber\\
{\rm Algorithm}    & : &  \pm 0.05~{\rm mags} \nonumber\\
{\rm Total}        & : &  \pm 0.10~{\rm mags} \nonumber\\
\end{eqnarray*}

\subsection{Andromeda IX}

Andromeda~IX ($0{\rm h} ~52{\rm m} ~52.8{\rm s}, +43^\circ ~12' ~0''$)
was  only   very  recently  discovered   by  \cite{zucker2004}  during
inspection of an SDSS scan of  the environs of M31. It is the faintest
galaxy so far discovered, with a surface brightness of \mbox{$\Sigma_V
\simeq 26.8$\,mags\,arcsec$^{-2}$.} and appears  to be very similar in
morphological properties  to the other dwarf  spheroidal companions of
M31.  Due to  its  recent discovery,  detailed  information about  its
stellar content has not yet been obtained.

\begin{figure*}
\begin{center}
\includegraphics[angle=270, width=13.5cm]{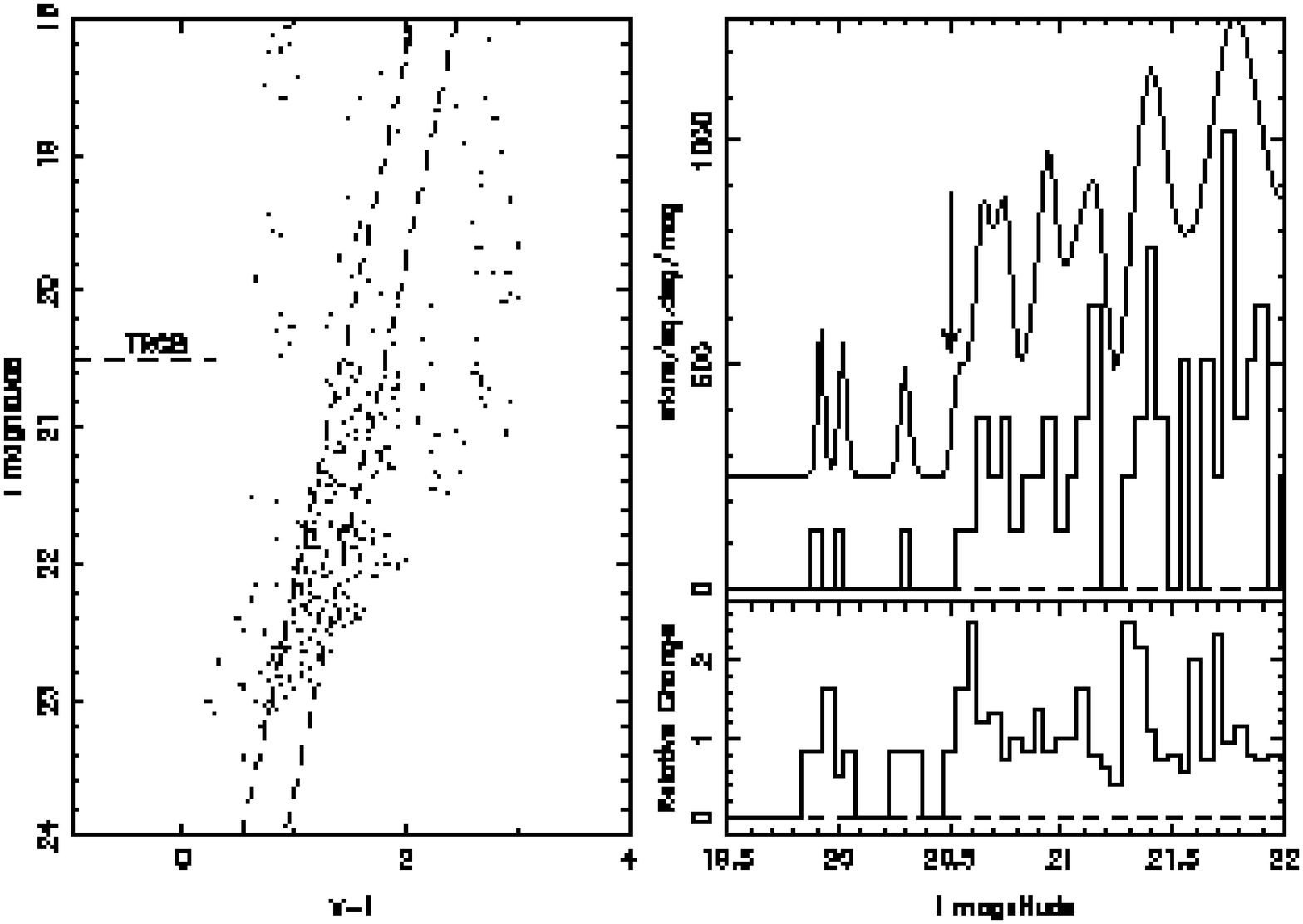}
\caption{And~IX CMD  (left panel), RGB luminosity  function and offset
LPD (upper right panel), and heuristic signal (lower right panel). The
TRGB is measured to lie at ${\rm I} = 20.50$~mags and is marked on the
CMD by a horizontal dashed line  and on the luminosity functions by an
arrow.}
\label{andix}
\end{center}
\end{figure*}

The  CMD  and   RGB  luminosity  function  for  And~IX   is  shown  in
Figure~\ref{andix}.   These  are  constructed  from all  stars  within
0\degg05 from the centre of this  galaxy. Their is an apparent lack of
foreground contamination present in this  field due to its small size,
and  comparison  of  the  And~IX  CMD  with  a  reference  CMD  offset
$10^{\prime}$ to the north-west shows there are virtually no M31 field
stars in the  region of the TRGB of And~IX. As  such, subtraction of a
reference field from the And~IX luminosity function does not alter our
result.  The onset of the TRGB in this galaxy is clear, and the result
of the application of our TRGB algorithm is shown below.

\begin{eqnarray*}
And~IX\\
{\rm I_{TRGB}}           & = & 20.50 \pm 0.03~{\rm mags} \nonumber\\
{\rm E\left(B - V\right)}& = & 0.077~{\rm mags}        \nonumber\\
{\rm \left[M/H\right]_{\alpha=0.0}}  & = & -1.5            \nonumber\\
{\rm \left[M/H\right]_{\alpha=0.3}}&=&-1.4\nonumber\\
{\rm M_I^{TRGB}}         & = & -4.065~{\rm mags}     \nonumber\\
{\rm \left(m - M\right)_o} & = & 24.42 \pm 0.07~{\rm mags}\nonumber\\
{\rm D_{And~IX}}        & = & 765 \pm 24~{\rm kpc}      \nonumber\\
\\
And~IX~Error~Budget&  & \nonumber \\
{\rm Photometry}-~~~rms   & : & \pm 0.02~{\rm mags} \nonumber\\
{\rm ~~~~~~~~~~}~-zeropt   & : & \pm 0.02~{\rm mags} \nonumber\\
{\rm Reddening}    & : & \pm 0.02~{\rm mags} \nonumber\\
{\rm M_I^{TRGB}}   & : &  \pm 0.05~{\rm mags} \nonumber\\
{\rm Algorithm}    & : &  \pm 0.03~{\rm mags} \nonumber\\
{\rm Total}        & : &  \pm 0.07~{\rm mags} \nonumber\\
\end{eqnarray*}

\subsection{Aquarius (DDO 210)}

Aquarius is located  at $20{\rm h} ~46{\rm m}  ~51.8{\rm s}, -12^\circ
~50' ~53''$ and is generally considered to be located at the periphery
of   the   Local  Group.    Discovered   in   the   late  fifties   by
\cite{vandenbergh1959}, an  in depth  study of its  stellar population
has recently been conducted  by \cite{lee1999}. This author found that
the galaxy  is relatively metal-poor, at ${\rm  [Fe/H]} \simeq -1.86$.
The central  regions have recently  seen enhanced star  formation, and
several  young stars  are observed.   An RGB  and AGB  are  also seen,
indicating the  presence of intermediate and  old stellar populations,
as would  be expected  for a dwarf  irregular galaxy.   Overall, there
appears to  be some evidence  for a disk  - halo split in  the stellar
content,  similar to WLM,  although this  has yet  to be  confirmed by
other observations.

\begin{figure*}
\begin{center}
\includegraphics[angle=270, width=13.5cm]{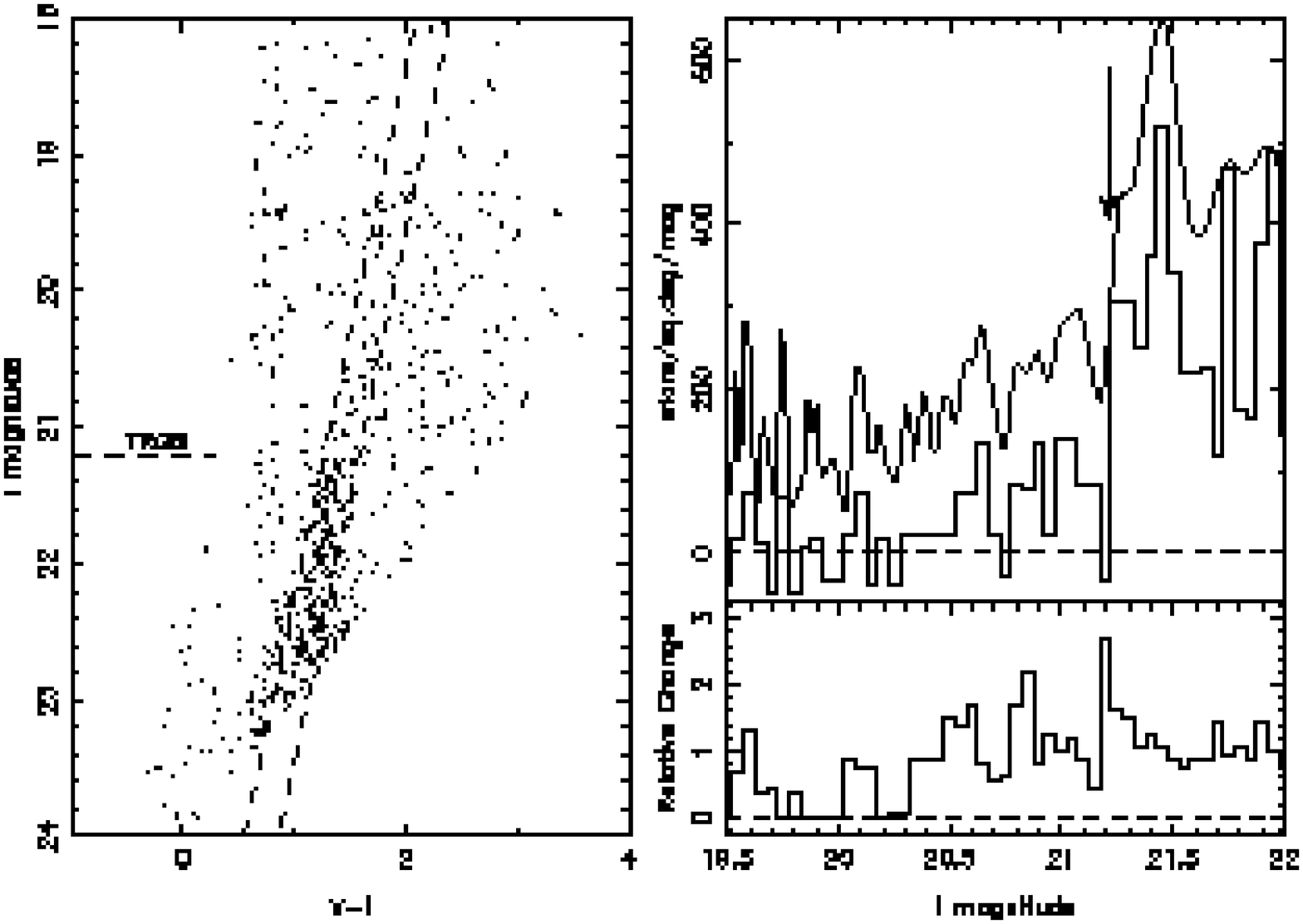}
\caption{Aquarius CMD (left panel), RGB luminosity function and offset
LPD (upper right panel), and heuristic signal (lower right panel). The
TRGB is measured to lie at ${\rm I} = 21.21$~mags and is marked on the
CMD by a horizontal dashed line  and on the luminosity functions by an
arrow.}
\label{aquarius}
\end{center}
\end{figure*}

The CMD  and RGB  luminosity functions for  Aquarius are  presented in
Figure~\ref{aquarius},  constructed from  the  innermost 0\degg075  of
this galaxy.   In addition to the RGB,  a blue plume of  stars is also
observed, demonstrating the existence  of a population of young stars,
in    agreement    with     \cite{lee1999}.     As    inspection    of
Figure~\ref{aquarius} shows, this is  the hardest galaxy in our sample
on which to perform our analysis. The difficulty is due to the feature
located  between ${\rm  I}  \sim 20.5  \rightarrow  21.2$ mags.   This
relatively bright excess could be the  onset of the RGB of Aquarius, a
bright AGB  population, or a  foreground population.  We rule  out the
latter option as  this feature seems to be  robust against any spatial
cuts and foreground corrections that we choose to apply.  Without more
information it  is difficult to distinguish between  the two remaining
possibilities.  Our  preferred interpretation is  that it is  a bright
AGB population - \cite{lee1999} have shown that we would expect to see
such a  feature.  The fainter excess  at $I >  21.2$~mags appears much
more populated and well-defined than  the proposed AGB feature, and so
we interpret this  as the RGB population.  Under  this hypothesis, the
TRGB algorithm  produces the results detailed  below. Additionally, as
was  the case  for LGS3,  a large  fraction of  the stars  in Aquarius
appear as metal  poor as our most metal  poor evolutionary track. This
has therefore been adopted as the typical metallicity of this system.

\begin{eqnarray*}
Aquarius\\
{\rm I_{TRGB}}           & = & 21.21 \pm 0.04~{\rm mags} \nonumber\\
{\rm E\left(B - V\right)}& = & 0.052~{\rm mags}        \nonumber\\
{\rm \left[M/H\right]_{\alpha=0.0}} & \lesssim & -2.3     \nonumber\\
{\rm \left[M/H\right]_{\alpha=0.3}}& \lesssim &-2.0\nonumber\\
{\rm M_I^{TRGB}}         & = & -4.04~{\rm mags}     \nonumber\\
{\rm \left(m - M\right)_o} & = & 25.15 \pm 0.08~{\rm mags}\nonumber\\
{\rm D_{Aquarius}}        & = & 1071 \pm 39~{\rm kpc}      \nonumber\\
\\
Aquarius~Error~Budget & &\nonumber \\
{\rm Photometry}-~~~rms  & : & \pm 0.04~{\rm mags} \nonumber\\
{\rm ~~~~~~~~~~}~-zeropt   & : & \pm 0.02~{\rm mags} \nonumber\\
{\rm Reddening}    & : & \pm 0.02~{\rm mags} \nonumber\\
{\rm M_I^{TRGB}}   & : &  \pm 0.05~{\rm mags} \nonumber\\
{\rm Algorithm}    & : &  \pm 0.04~{\rm mags} \nonumber\\
{\rm Total}        & : &  \pm 0.08~{\rm mags} \nonumber\\
\end{eqnarray*}

\section{Summary}
 
\begin{figure*}
\begin{center}
\includegraphics[angle=270, width=17cm]{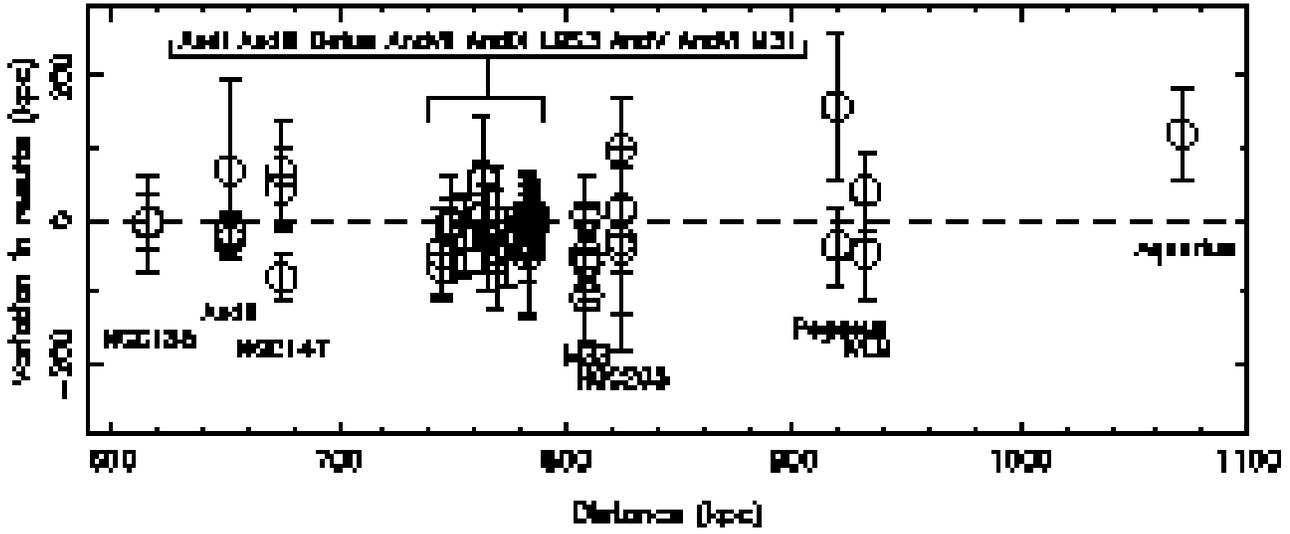}
\caption{The distances  we derive for  the galaxies listed  in Table~2
compared to  the selection of  previous estimates also listed  in that
Table.   The x-axis  is  the distance  we  derive; the  y-axis is  the
difference between  our result and  the previous estimates,  such that
points lying  below the  dashed line are  previous estimates  that are
larger  than those  presented here.  Given the spread in distance values
for any given galaxy and the relatively large error bars, there are no
obvious biases or trends in  our distance estimates.  In general, they
are seen to be in good agreement with previous work.}
\label{comparison}
\end{center}
\end{figure*}

Table~1 lists the  galaxies analysed as part of  this study, including
those from Paper~I, along with their positions, the adopted values for
the reddening, the median metallicities as determined by the colour of
the red giant  branch and the adopted value  of $M_I^{TRGB}$ using the
modified  calibration  of  \cite{bellazzini2004}.   Table~2  list  the
measured position of  the TRGB, the distance modulus  and distance for
each galaxy.   The three galaxies originally analysed  in Paper~I have
had their distances recalcuated  to correct for the slight metallicity
dependancy that was not previously addressed in Paper~I. Additionally,
the TRGB location  in M33 has been rederived  using a local foreground
correction  (in  the same  way  as for  M31)  and  using only  objects
that lie within $2\,-\,\sigma$ of the stellar locus. Also  included for
comparision  are a selection  of previous  distance estimates  to each
system from the literature.  Figure~\ref{comparison} shows a graphical
comparison  between our  results and  these earlier  measurements.  In
general, good agreement is  observed and no obvious systematic offsets
or trends  are visible.  For  those cases where Cepheid  distances are
available       (M31:      \citealt{joshi2003,freedman1990};      M33:
\citealt{lee2002}) our results match  these estimates to within better
than a few tens of kiloparsecs.

For a  few of the fainter objects  in our sample (eg.  And~IX) we have
relatively few  bright RGB stars available  to us, due to  the lack of
such  stars in  these  systems. The  distance measurements  implicitly
assume that the brightest RGB  stars in the system are good indicators
of the  actual position of the TRGB.  For the case of  And~IX, the RGB
luminosity  function  shows  a  steep  rise and  is  clearly  defined,
compensating  in  part for  the  fewer  stars  available to  us.  Thus
relatively  accurate  TRGB  measurements  can be  obtained  for  these
systems,  even although the  number of  bright RGB  stars is  far less
than, say, NGC~205 or M31.

We conclude with a list of  some of the main results evident from this
study:

1. The distance to the  newly discovered dwarf spheroidal companion to
   M31, Andromeda~IX, is  measured to be $765 \pm  25$~kpc. This is in
   good  agreement to  the distance  calculated  by \cite{zucker2004}.
   Assuming   And~IX   does  not   have   a  strong   $\alpha$-element
   enhancement, its metallicity is measured to be \mbox{[Fe/H] $\simeq
   -1.5$}. Figure~4 of \cite{caldwell1999} shows an empirical relation
   that  is   found  to   exist  between  integrated   luminosity  and
   metallicity  for  dwarf  spheroidal  galaxies,  such  that  fainter
   galaxies   have  lower   metallicity.    \cite{zucker2004}  measure
   $M_{tot,V} \simeq -8.3$\,mags which  would imply that And~IX should
   have  a  metallicity of  approximately  \mbox{[Fe/H] $\sim  -2.2$.}
   And~IX  thus appears  to  be more  metal-rich  than this  empirical
   relation would suggest.

2. And~V is measured to have \mbox{[Fe/H] = -1.6}, again assuming no
   $\alpha$-enhancement.  This disagrees with the recent measurement
   by \cite{davidge2002} of \mbox{[Fe/H] = -2.2}.  It does agree,
   however, with the earlier measurement by \cite{armandroff1998}.
   Both this latter measurement and our measurement indicate that,
   like And~IX, And~V does not lie on the expected relation between
   integrated luminosity and metallicity. These results suggest that
   And~I, II, III, V, VI, VII \& IX all have comparable metallicities
   in the range \mbox{$-1.7 \le$ [Fe/H] $\le 1.4$}, with an estimated
   accuracy on the measurements of $\pm\,\sim 0.1\,-\,0.2$\,dex.

3. Aquarius is confirmed to lie $\sim 1$~Mpc from the Milky Way. Early
   estimates  of   its  distance  had  placed  it   well  outside  the
   zero-velocity  surface  of  the  Local Group  until  \cite{lee1999}
   showed that it was substantially  closer than this. We note however
   that the distance to Aquarius  is the least certain from this study
   for the reasons discussed in Section~3.12. It could potentially lie
   closer to us than we have derived here.

4. Pegasus is measured  to lie at $919 \pm  30$~kpc.  This measurement
   agrees with earlier work done by \cite{aparicio1994}. A more recent
   study  by  \cite{gallagher1998}  places  this object  some  160~kpc
   closer -  this distance  is required  by their study  so as  to fit
   self-consistent  stellar population  models based  upon  the Geneva
   stellar evolutionary  tracks. \cite{gallagher1998} also  require to
   adopt  a  larger reddenning  value  than  is  normally assumed  for
   Pegasus  (${\rm E(B  -  V)} \simeq  0.14$)  in order  to match  the
   colours  of  the  main  sequence.   Even by  adopting  this  larger
   reddening value, the TRGB  distance appears incompatible with their
   measurement of the distance.

5. NGC~185  and NGC~147 have  previously been  suspected of  forming a
   binary  system  due  to   their  small  angular  seperation  ($\sim
   1^\circ$)     and    similar     line     of    sight     distances
   (\citealt{vandenbergh1998}). The distances that we derive for these
   objects suggest  that they are likely to  be gravitationally bound,
   separated in  line of sight  by only $\sim 60$\,kpc.   $1^\circ$ at
   the distance of  this system corresponds to $\sim  11$\,kpc, and so
   their physical  seperation is  also of order  60 kpc,  although the
   uncertainties  in each of  their distances  means that  the precise
   value may be substantially different to this.

The 17 Local Group galaxies listed in Tables~1~\&~2 were observed with
the same telescope, instrument and filters for similar exposure times
and the resulting data were reduced using the same pipeline
processing.  An identical analysis was then conducted on each to
calculate the position of the TRGB, its median metallicity and its
distance.  The use of a Wide Field Camera has maximised the number of
stars observed per pointing.  This reduces the effect of Poisson noise
on our measurements and ensures that the luminosity function that we
derive is as accurate a representation of the intrinsic luminosity
function of the galaxy as is possible in the region of the tip. The
resulting set of metallicity and distance estimates have thus had
systematic uncertainties minimised and are therefore ideal for any
study of Local Group galaxy properties.

\section * {Acknowledgements}
We  are  once again  grateful  to  the  referee, Barry  Madore,  whose
comments greatly helped to increase  the clarity of this work.  Thanks
also go to Blair Conn and  Jonathan Irwin for helping take much of the
data presented in this paper. AWM  thanks Gary Da Costa for a valuable
discussion regarding  these results. We  would also like to  thank Ata
Sarajedini, Glenn Tiede and Michael  Barker for sharing their M33 data
with us which lead us to  refine our distance estimate to this galaxy.
The research of AMNF has been supported by a Marie Curie Fellowship of
the European Community  under contract number HPMF-CT-2002-01758. This
work was based on observations made with the Isaac Newton Telescope on
the  Island of  La Palma  by  the Isaac  Newton Group  in the  Spanish
Observatorio del Roque de los Muchachos of the Institutode Astrofisica
de Canarias.

\bibliographystyle{apj}
\bibliography{/home/alan/morespace/latex/papers/references}

\begin{table*}
\begin{minipage}{0.90285\textwidth}
\begin{tabular*}{0.90285\textwidth}{|l|cc|c|c|c|c|c|c|}
\hline
              &             &             &          & \multicolumn{2}{|c|}{[M/H]}  & \multicolumn{3}{|c|}{$M_I^{TRGB}$}\\
\cline{5-9}   & \raisebox{1.5ex}{RA (J2000)} & \raisebox{1.5ex}{Dec (J2000)} & \raisebox{1.5ex}{E(B - V)} & $\alpha=0.0$& $\alpha=0.3$   & $\alpha=0.0$&$\alpha=0.3$&Adopted\\
\hline
          &                                   &                     &      &      &       &       &      & \\
M31       & $00{\rm h} 42{\rm m} 44.3{\rm s}$ & $+41^\circ 16' 09''$& 0.060& -0.6 & -0.5  & --- & ---& -4.05 \footnote{Calculated using the representative metallicity of the RGB stars contained within the strip used in the TRGB measurement. Due to the width of the RGB in M31, M33 and NGC~205,  this does not correspond to the median metallicity of the RGB in these cases (see Figure~\ref{m31} for M31, Figure~5 in Paper~I for M33, and Figure~\ref{ngc205} for NGC~205). In each of these cases the representaative metallicity is [M/H]$ \sim -1$.}\\
          &                                   &                     &      &      &       &       &      & \\
M33       & $01{\rm h} 33{\rm m} 50.9{\rm s}$ & $+30^\circ 39' 36''$& 0.042& -0.8 & -0.7  & --- & --- & -4.05$^{a}$\\
          &                                   &                     &      &      &       &       &      & \\
NGC~205   &$0{\rm h} 40{\rm  m} 22.1{\rm  s}$ & $+41^\circ 41' 07''$& 0.062& -0.8 & -0.7  & --- & --- & -4.05$^{a}$\\
          &                                   &                     &      &      &       &       &      & \\
NGC~185   & $00{\rm h} 38{\rm m} 58.0{\rm s}$ & $+48^\circ 20' 15''$& 0.179& -1.2 & -1.1  & -4.07 & -4.06 & -4.065\\
          &                                   &                     &      &      &       &       &      & \\
NGC~147   & $00{\rm h} 33{\rm m} 12.1{\rm s}$ & $+48^\circ 30' 32''$& 0.175& -1.1 & -1.0  & -4.06 & -4.05 &-4.055 \\
          &                                   &                     &      &      &       &       &      & \\
Pegasus   & $23{\rm h} 28{\rm m} 36.2{\rm s}$ & $+14^\circ 44' 35''$& 0.064& -1.4 & -1.2  & -4.07 & -4.07 & -4.07\\
          &                                   &                     &      &      &       &       &      & \\
WLM       & $00{\rm h} 01{\rm m} 58.1{\rm s}$ & $-15^\circ 27' 39''$& 0.035& -1.5 & -1.4  & -4.06 & -4.07 & -4.065\\
          &                                   &                     &      &      &       &       &      & \\
LGS3      & $01{\rm h} 03{\rm m} 52.9{\rm s}$ & $+21^\circ 53' 05''$& 0.042&$\lesssim -2.3$ &$\lesssim -2.0$& -4.04 & -4.04 &-4.04\\
          &                                   &                     &      &      &       &       &      & \\
Cetus     & $00{\rm h} 26{\rm m} 11.0{\rm s}$ & $-11^\circ 02' 40''$& 0.029& -1.6 & -1.5  & -4.05 & -4.06 &-4.055\\
          &                                   &                     &      &      &       &       &      & \\
And~I     & $00{\rm h} 45{\rm m} 39.8{\rm s}$ & $+38^\circ 02' 28''$& 0.056& -1.4 & -1.3  & -4.07 & -4.07 &-4.07\\
          &                                   &                     &      &      &       &       &      & \\
And~II    & $01{\rm h} 16{\rm m} 29.8{\rm s}$ & $+33^\circ 25' 09''$& 0.063& -1.5 & -1.4  & -4.06 & -4.07 &-4.065\\
          &                                   &                     &      &      &       &       &      & \\
And~III   & $00{\rm h} 35{\rm m} 33.8{\rm s}$ & $+36^\circ 29' 52''$& 0.058& -1.7 & -1.6  & -4.04 & -4.05 &-4.045\\
          &                                   &                     &      &      &       &       &      & \\
And~V     & $01{\rm h} 10{\rm m} 17.1{\rm s}$ & $+47^\circ 37' 41''$& 0.124& -1.6 & -1.5  & -4.05 & -4.06 &-4.055\\
          &                                   &                     &      &      &       &       &      & \\
And~VI    & $23{\rm h} 51{\rm m} 46.3{\rm s}$ & $+24^\circ 34' 57''$& 0.065& -1.5 & -1.3  & -4.06 & -4.07 &-4.065\\
          &                                   &                     &      &      &       &       &      & \\
And~VII   & $23{\rm h} 26{\rm m} 31.0{\rm s}$ & $+50^\circ 41' 31''$& 0.199& -1.4 & -1.3  & -4.07 & -4.07 &-4.07\\
          &                                   &                     &      &      &       &       &      & \\
And~IX    & $0{\rm h} 52{\rm m} 52.8{\rm s}$  & $+43^\circ 12' 00''$& 0.076& -1.5 & -1.4  & -4.06 & -4.07 &-4.065\\
          &                                   &                     &      &      &       &       &      & \\
Aquarius  & $20{\rm h} 46{\rm m} 51.8{\rm s}$ & $-12^\circ 50' 53''$& 0.052&$\lesssim -2.3$ &$\lesssim -2.0$& -4.04 & -4.04 &-4.04\\
          &                                   &                     &      &      &       &       &      & \\
\hline
\end{tabular*}
\caption{The  position,  adopted  reddening,  median  metallicity  and
adopted  value  of  $M_I^{TRGB}$  for  each of  the  galaxies  in  our
dataset. The reddening  values have been taken from  the maps produced
by Schlegel et  al. (1998) and are related to the  extinction in the I
band  via $A_I  =  1.94 {\rm  E\left(  B -  V  \right)}$.  The  median
metallicity has been calculated for each system using the evolutionary
tracks produced  by Vandenbergh et  al. (2000) and is  calculated such
that  it is  consistent with  the  TRGB measurement.   Two values  are
quoted as we consider  both $\alpha=0.0$ and $\alpha=0.3$. The implied
value  of $M_I^{TRGB}$  has been  calculated for  each case  using the
modified calibration  of Bellazzini et  al. (2004) (see  Section 2.4).
There is  little difference  in $M_I^{TRGB}$ between  the $\alpha=0.0$
and $\alpha=0.3$ cases.  LGS3 and  Aquarius are at least as metal poor
as our most metal poor evolutionary track. The three galaxies analysed
in  Paper~I are  included  in  this Table,  and  their final  distance
estimates have been revised accordingly.}
\end{minipage}
\end{table*}




\begin{table*}
\begin{minipage}{\textwidth}
\begin{tabular*}{\textwidth}{@{\extracolsep{\fill}}|l|cc|ccc|cc|}
\hline \vspace{-0.35cm}\\
    &                   &                  &       &        &     &                          &                \\
    & ${\rm I}_{TRGB}$  & $\left( m - M \right)_\circ$ & D (kpc)&& $\Delta\,D$ (kpc) & Previous Estimates & References\vspace{-0.25cm} \\
    &                   &                  &       &        &     &                          &                \\
\hline \vspace{-0.45cm}\\
    &                   &                  &       &        &     &                          &                \\
M31 &$ 20.54 \pm 0.03 $ & $24.47 \pm 0.07$ & 785   &$\pm$   & 25  & $794 \pm 37$ kpc         &\cite{brown2004}\\
    &                   &                  &       &        &     & $791 \pm 40$ kpc         &\cite{joshi2003}\\
    &                   &                  &       &        &     & $783 \pm 43$ kpc         &\cite{durrell2001}\\
    &                   &                  &       &        &     & $773 \pm 36$ kpc         &\cite{freedman1990}\\
    &                   &                  &       &        &     &                          &                \\
M33 & $20.57 \pm 0.03 $ & $24.54 \pm 0.06$ & 809   &$\pm$    & 24  & $855 \pm 60$ kpc        &\cite{galleti2004} \\
    &                   &                  &       &        &  & $867 \pm 28$ kpc            &\cite{tiede2004} \\
    &                   &                  &       &        &  & $802 \pm 51$ kpc            &\cite{lee2002} \\
    &                   &                  &       &        &  & $916 \pm 55$ kpc            &\cite{kim2002} \\
    &                   &                  &       &        &  &                             &                \\
NGC~205   & $20.65 \pm 0.03$ & $ 24.58 \pm 0.07$   & 824 &$\pm$& 27 & $809 \pm 80$ kpc       &\cite{salaris1998} \\
    &                   &                  &       &        &  & $851 \pm 98$ kpc            &\cite{saha1992} \\
                                                       &      &                                         &          &       &                    & $863 \pm 139$ kpc &\cite{ciardullo1989} \\
                                                       &       &                                      &          &       &                    & $724 \pm 67$ kpc &\cite{mould1984} \\
    &                   &                  &       &          & &                        &                \\
NGC~185   & $20.23 \pm 0.03$ & $23.95 \pm 0.09$  & 616 &$\pm$& 26 & $617 \pm 28$ kpc      &\cite{martinezdelgado1998} \\
                                                       &         &                                      &          &       &                    & $620 \pm 60$ kpc &\cite{lee1993b} \\
    &                   &                  &       &          &   &                      &                \\
NGC~147   & $20.43 \pm 0.04$ & $24.15 \pm 0.09$ & 675 &$\pm$& 27 & $755 \pm 17$ kpc        &\cite{han1997} \\
                                                       &           &                                    &          &       &                    &$608 \pm 70$ kpc    &\cite{saha1990a}\\
                                               &                    &                    &          &       &                    &$630 \pm 50$ kpc    &\cite{mould1983}\\
    &                   &                  &       &          &      &                   &                \\
Pegasus   & $20.87 \pm 0.03$ & $24.82 \pm 0.07$ & 919 &$\pm$& 30 & $760\pm 100$ kpc      &\cite{gallagher1998} \\
                                                       &              &                                &          &       &                    & $955 \pm 44$ kpc      &\cite{aparicio1994} \\
                                                       &               &                                &          &       &                    &$1.7 \pm 0.23$ Mpc    &\cite{hoessel1982}\\
    &                   &                  &       &          &         &                &                \\
WLM       & $20.85 \pm 0.05$ & $24.85 \pm 0.08$ & 932 &$\pm$& 33 & $977 \pm 58$ kpc      &\cite{rejkuba2000}\\
                                                       &                 &                             &          &       &                    & $891 \pm 41$ kpc      &\cite{minniti1997}\\
    &                   &                  &       &          &           &              &                \\
LGS3      & $20.47 \pm 0.03$  & $24.43 \pm 0.07$  & 769 &$\pm$& 23 & $770 \pm 70 $ kpc   &\cite{aparicio1997}\\
                                                       &                   &                           &          &       &                    & $810 \pm 80$ kpc      &\cite{lee1995} \\
    &                   &                  &       &          &             &            &                \\
Cetus     & $20.39 \pm 0.03$ & $24.39 \pm 0.07$   & 755 &$\pm$& 23 & $780 \pm 50$            &\cite{sarajedini2002} \\
                                                       &                     &                        &          &       &                    & $775 \pm 50$            &\cite{whiting1999} \\
    &                   &                  &       &          &               &          &                \\
And~I     & $20.40 \pm 0.03$ & $24.36 \pm 0.07$ & $745$ &$\pm$& 24 & $810 \pm 30$ kpc      &\cite{dacosta1996} \\
                                                       &                       &                        &          &       &                    & $790 \pm 60$ kpc      &\cite{mould1990} \\
    &                   &                  &       &          &                 &        &                \\
And~II   & $20.13 \pm 0.02$ & $24.07 \pm 0.06$ & 652 &$\pm$& 18 & $665 \pm 20$ kpc      &\cite{pritzl2004} \\
                                                       &                         &                      &          &       &                    & $680 \pm 20$ kpc &\cite{dacosta2000} \\
                                                       &                          &                     &          &       &                    & $583^{+124}_{-103}$ kpc &\cite{koenig1993} \\
    &                   &                  &       &          &                    &     &                \\
And~III   & $20.44 \pm 0.04$ & $24.37 \pm 0.07$  & 749 &$\pm$& 24 & $752 \pm 21$ kpc        &\cite{dacosta2002} \\
                                                       &                            &                   &          &       &                     & $758 \pm 70$ kpc         &\cite{armandroff1993}\\
    &                   &                  &       &          &                      &   &                \\
And~V     & $20.63 \pm 0.04$& $24.44 \pm 0.08$  & 774 &$\pm$& 28 & $810 \pm 45$ kpc        &\cite{armandroff1998} \\
    &                   &                  &       &          &                       &  &                \\
And~VI    & $20.53 \pm 0.04$ & $24.47 \pm 0.07$ & 783 &$\pm$& 25 & $815 \pm 25$ kpc         &\cite{pritzl2002} \\
                                                      &                                &               &          &       &                     & $830 \pm 80$ kpc         &\cite{grebel1999a}\\
                                                       &                                &               &          &       &                     & $820 \pm 94$ kpc         &\cite{tikhonov1999}\\
                                                       &                                 &              &          &       &                    & $775 \pm 35$ kpc        &\cite{armandroff1999} \\
                                                       &                                  &             &          &       &                   & $794 \pm 73$ kpc        &\cite{hopp1999}  \\
    &                   &                  &       &          &                         &  &              \\
And~VII   & $20.73 \pm 0.05$ & $24.41 \pm 0.10$  & 763 &$\pm$& 35 & $760 \pm 70$ kpc        &\cite{grebel1999a} \\
                                                       &                                    &          &          &       &                    & $708 \pm 81$ kpc &\cite{tikhonov1999}\\
    &                   &                  &       &          &                         &    &            \\
And~IX   & $20.50 \pm 0.03$ & $24.42 \pm 0.07$  & 765 &$\pm$& 24 & $790 \pm 70$ kpc        &\cite{zucker2004} \\      
    &                   &                  &       &          &                         &      &          \\
Aquarius  & $21.21 \pm 0.04$ & $25.15 \pm 0.08$   & 1071 & $\pm$ & 39 & $950 \pm 50$ kpc            &\cite{lee1999} \\                                                   &                & & & & &$\sim 4$ Mpc &\cite{greggio1993}\vspace{-0.17cm}\\     
&                   &                  &       &          &                         &    &            \\
\hline
\end{tabular*}
\caption{The measured value of  the TRGB, implied distance modulus and
distance to  each galaxy analysed, along with  previous estimates. The
latter are representative only and do not form an exhaustive list. The
distance  estimates to  the  galaxies analysed  in  Paper~I have  been
revised in  accordance with the  methods presented in this  paper.  In
addition, the TRGB  location for M33 has been  rederived using a local
foreground  correction in  the  same way  as  for M31  and using  only
objects lying within $2 - \sigma$ of the stellar locus.}
\end{minipage}
\end{table*}



\end{document}